\documentclass[preprints,article,accept,moreauthors,pdftex]{mdpi}
\firstpage{1}
\pubvolume{xx}
\issuenum{1}
\articlenumber{5}
\pubyear{2020}
\copyrightyear{2020}
\usepackage{amssymb}
\usepackage{graphics}
\usepackage{amsmath}
\usepackage{amsfonts}
\usepackage{color}

\definecolor{darkblue}{rgb}{0,0,.7}

\newcommand{\be}{\begin{eqnarray}}
\newcommand{\ee}{\end{eqnarray}}

\newcommand{\ide}{1\hspace{-0.8mm}{\rm I}}

\newcommand{\ket}[1]{\mbox{$\mid \!#1\rangle$}}

\history{Received: date; Accepted: date; Published: date}
\Title{From R\'{e}nyi Entropy Power to Information Scan of Quantum States}

\Author{Petr Jizba $^{1,\ddagger}$*\orcidA{}, Jacob Dunningham $^{2,\ddagger}$\orcidB{}  and Martin Prok\v{s} $^{1,\ddagger}$}

\AuthorNames{Petr Jizba, Jacob Dunningham and Martin Prok\v{s}}

\address{%
$^{1}$ \quad FNSPE, Czech Technical University in Prague, B\v{r}ehov\'{a} 7, 115 19 Praha 1, Czech Republic; \\
$^{2}$ \quad Department of Physics and Astronomy, University of Sussex, Brighton BN1 9QH, United Kingdom;
}
\corres{Correspondence: p.jizba@fjfi.cvut.cz; Tel.: +420 775-317-307 (P.J.)}

\secondnote{These authors contributed equally to this work.}
\abstract{In the estimation theory context, we generalize the notion of Shannon's entropy power
to the R\'{e}nyi-entropy setting. This not only allows to find new estimation inequalities, such as the
R\'{e}nyi-entropy based De Bruijn identity, isoperimetric inequality or Stam inequality, but
it also provides a convenient technical framework for the derivation of a one-parameter family of R\'{e}nyi-entropy-power-based
quantum-mechanical uncertainty relations.
To put more flesh on the bones, we use the R\'{e}nyi entropy power obtained to show how the information
probability distribution associated with a quantum state can be reconstructed
in a process that is akin to quantum-state tomography. We illustrate the inner workings of this with
the so-called ``cat states'', which are of fundamental interest and practical use in schemes such as quantum metrology.
Salient issues, including the extension of the notion of entropy power to Tsallis entropy and
ensuing implications in estimation theory are also briefly discussed.
}

\keyword{R\'{e}nyi entropy; Tsallis entropy; Entropic uncertainty relations; Quantum metrology.}

\begin{document}
%%%%%%%%%%%%%%%%%%%%%%%%%%%%%%%%%%%%%%%%%%

%%%%%%%%%%%%%%%%%%%%%%%%%%%%%%%%%%%%%%%%%%
%\setcounter{section}{-1} %% Remove this when starting to work on the template.
\section{Introduction~\label{sec1}}
%%%%%%%%%%%%%%%%%%%%%%%%%%%%%%%%%%%%%%%%%

The notion of entropy is undoubtedly one of the most important concepts in modern science.
Very few other concepts can compete with it in respect to the number of attempts to clarify
its theoretical and philosophical meaning. Originally, the notion of entropy stemmed from thermodynamics,
where it was developed to quantify the annoying inefficiency of steam engines.
It then transmuted into a description of the amount of disorder or complexity in physical systems.
Though many such attempts were initially closely connected with the statistical interpretation
of the phenomenon of heat, in the course of time, they expanded their scope
far beyond their original incentives. Along those lines
several approaches have been developed in attempts to quantify and qualify the entropy paradigm. These have been
formulated largely independently and with different applications and goals in
mind. For instance, in {\em statistical physics}, the entropy counts the number of distinct microstates compatible with a given
macrostate, in {\em mathematical statistics}, it corresponds to the inference functional for an updating procedure, and in {\em information theory}, it determines a limit on the shortest attainable encoding scheme.

Particularly distinct among these are the information-theoretic entropies (ITEs). This is not only because they
discern themselves through their firm operational prescriptions in terms of coding theorems and communication protocols~\cite{Shannon48,Shannon49,Feinstein58,Campbell65,Bercher09},
but because they also offer an intuitive  measure of disorder phrased in terms of missing information about a system. Apart from innate issues in communication theory,
ITEs have also proved to be indispensable tools in other branches of science. Typical examples are provided by chaotic dynamical systems and multifractals (see e.g., \cite{Thurner18} and citations therein).
Fully developed turbulence, earthquake analysis and generalized dimensions of strange attractors provide further examples~\cite{Tsallis-book}. An especially important arena for ITEs in the past two
decades has been quantum mechanics (QM) with applications ranging from quantum estimation and coding theory to quantum entanglement. The catalyst has been an infusion of new
ideas from (quantum) information theory~\cite{Maasen:88,Birulla:07,Birula,JD:16},
functional analysis~\cite{Ozawa:16,Dang}, condensed matter theory~\cite{Zeng18,Melcher19} and cosmology~\cite{Ryu06,Eisert10}. On the experimental front, the use of ITEs has been stimulated
not only by new high-precision instrumentation~\cite{Pikovski:12,Marin:13} but also by, e.g. recent advances in stochastic  thermodynamics~\cite{An14,Campisi11} or observed violations of Heisenberg's error-disturbance
uncertainty relations~\cite{Ozawa:12,Ozawa:13a,Ozawa:13b,Dressel,Busch} .

In his seminal 1948 paper, Shannon laid down the foundations of modern information theory~\cite{Shannon48}. He was also instrumental in pointing out that, in contrast with discrete signals or messages where information is quantified by (Shannon's) entropy, the cases with continuous variables are less satisfactory. The continuous version of Shannon's entropy (SE) --- the so-called differential entropy, may take negative values~\cite{Shannon48,Jizba2004}, and so it does not have the same status as its discrete-variable counterpart. To solve a number of information-theoretic problems related to continuous cases Shannon shifted the emphasis from the differential entropy to yet another object --- entropy power (EP). The EP describes the variance of a would-be Gaussian random variable with the same differential entropy as the random variable under investigation. EP was used by Shannon~\cite{Shannon48,Shannon49} to bound the capacity of non-Gaussian additive noise channels. Since then, the EP has proved to be essential in a number of applications ranging from interference
channels to secrecy capacity~\cite{5,6,7,8,9}. It has also led to new advances in information parametric statistics~\cite{Costa1985,frieden} and network information theory~\cite{Courtade}. Apart from its significant role in information theory, the EP has found wide use in pure mathematics, namely in the theory of inequalities~\cite{Courtade} and mathematical statistics and estimation theory~\cite{Barron:86}.

Recent developments in information theory~\cite{entropy_spec_issue}, quantum theory~\cite{biro,biro_II}, and complex dynamical systems in particular~\cite{Thurner18,thurner_II,thurner_IIa} have brought about the need for a further extension of the concept of ITE  beyond Shannon's conventional type. Consequently, numerous generalizations have started to proliferate in the literature ranging from additive entropies~\cite{Jizba2004,Burg} through a rich class of non-additive entropies~\cite{tsallis2,havrda,frank,sharma,korbel,arimitsu} to more exotic types of entropies~\cite{Vos}. Particularly prominent among such generalizations are ITEs of R\'{e}nyi and Tsallis, which both belong to a broader class of so-called Uffink entropic functionals~\cite{uffink,JK-PRL:19}. Both  R\'{e}nyi entropy (RE) and Tsalli entropy (TE) represent one-parameter families of deformations of Shannon's entropy.  An important point related to the RE is that the RE is not just a theoretical construct, but it has a firm operational meaning in terms of various coding theorems~\cite{Campbell65,Bercher09}. Consequently,  REs alongside with their associated R\'{e}nyi entropy powers (REPs) are, in principle, experimentally accessible~\cite{Cambpel,JizbaPRE,Elben}. That is indeed the case in specific quantum protocols~\cite{Bacco,Muller,Coles}. In addition, REPs of various orders are often used as convenient measures of entanglement --- e.g., REP of order $2$, i.e., $N_2$ represents {\em tangle} $\tau$ (with $\sqrt{\tau}$ being {\em concurrence})~\cite{Minter}, $N_{1/2}$ is related to both {\em fidelity} $F$ and {\em robustness}  $R$ of a pure state~\cite{Vidal}, $N_{\infty}$ quantifies the Bures distance to the closest separable pure state~\cite{Bengtsson}, etc. Even though our main focus here will be on REs and REPs since they are more pertinent in information theory, we will include some discussion related to Tsallis entropy powers at the end of this paper.

The aim of this paper is twofold. First, we wish to appropriately extend the notion of SE-based EP to the RE setting.
In contrast to our earlier works on the topic~\cite{JD:16,JDJ}, we will do it now by framing REP in the context of  RE-based
estimation theory. This will be done by judiciously generalizing such key notions as the De Bruijn identity, isoperimetric inequality (and ensuing Cram\'{e}r--Rao inequality) and Stam inequality.  In contrast to other similar works on the subject, our approach is distinct in the
in two key respects; a) we consistently use the notion of escort distribution and escort score vector in setting up the generalized De Bruijn identity and Fisher information matrix and b) R\'{e}nyi EP is related to variance of the reference Gaussian distribution rather than Tsallis type distribution.  As a byproduct  we derive
within such a generalized estimation theory framework the R\'{e}nyi-EP-based quantum uncertainty relations (REPUR) of Schr\"{o}dinger--Roberston type. The REPUR obtained
coincides with our earlier result~\cite{JD:16} that was obtained in a very different context by means of the Beckner--Babenko theorem. This in turn serves as a consistency check of the proposed generalized estimation theory.
Second, we identify interesting new playgrounds for the R\'{e}nyi EPs obtained. In particular, we asked ourselves a question: assuming one is able in specific quantum protocols
to measure R\'{e}nyi EPs of various orders, how does this constrain the underlying quantum state distribution?
To answer this question we invoke the concept of the {\em information distribution} associated
with a given quantum state. The latter contains  a complete ``information scan'' of the underlying state distribution.
%The advantage of these REPURs  over the conventional variance-based
%Robertson--Schr\"{o}dinger relation is that the right-hand side of the inequality is always
%independent of the considered quantum state and, in fact, it is a universal number
%for all EP orders.
We set up a reconstruction method based on Hausdorff's moment problem~\cite{Widder46}
to show explicitly how the information probability distribution associated with a given quantum state
can be numerically reconstructed from EPs in a process that is analogous to a quantum-state tomography.

The paper is structured as follows. In Sec.~2
%with a discussion briefly motivating the use of
we introduce the concept of R\'{e}nyi's EP. With quantum metrology applications in mind, we discuss this in the framework of estimation theory.
First,  we duly generalize the notion of Fisher information (FI) by using a R\'{e}nyi entropy
version of De~Bruijn's identity.  In this connection, we the emphasize role of the so-called {\em escort distribution}, which appears naturally in the definition of
higher-order {\em score functions}. Second, we prove the RE-based isoperimetric inequality and ensuing Cram\'{e}r--Rao inequality and find how
the knowledge of Fisher information matrix restricts possible values of R\'{e}nyi's EP.
Finally, we further illuminate the role of R\'{e}nyi's EP by deriving (through the Stam inequality)
R\'{e}nyi's EP-based quantum uncertainty relations for conjugate observables.
%
%and the ensuing generalized Cram\'{e}r--Rao's inequalities.
To flesh this out, the second part of the paper is devoted to the development of the use
of R\'{e}nyi EPs to extract the quantum state from incomplete data. This is of particular interest in various quantum metrology protocols.
To this end, we introduce in Sec.~3 the concepts of  information distribution and in Sec.~4 we show
how cumulants of the information distribution can be obtained from knowledge of the EPs. With the cumulants at hand, one can
reconstruct the underlying information distribution in a process which we call  an {\em information scan}.
Details of how one could explicitly realize such an information scan for quantum state PDFs is provided in Sec.~5.
There we employ generalized versions of  Gram--Charlier A and the Edgeworth expansion.
In Sec.~6 we illustrate the inner workings of the information scan using the example of a so-called {\em cat state}. This state is of interest in applications of quantum physics such as quantum-enhanced metrology, which is concerned with the optimal extraction of information from measurements subject to quantum mechanical effects. The cat state we consider is a superposition of the vacuum state and a coherent state of the electromagnetic field; two cases are studied comprising of different probabilistic weightings of the superposition state corresponding to \textit{balanced} and \textit{unbalanced} cat states. Sec.~7 is  dedicated to EPs based on
Tsallis entropy. In particular, we show that R\'{e}nyi and Tsallis EPs coincide with each other. This, in turn allows us to phrase various estimation theory inequalities in terms of TE.
In Sec.~7 we end with conclusions. For the reader's convenience, we relegate some technical issues concerning the generalized De~Bruijn identity and associated isoperimetric and Stam inequalities to three appendices.

%%%%%%%%%%%%%%%%%%%%%%%%%%%%%%%%%%%%%%%%%%%%%%%%%%%%%%%%%%%%%%%%%%%%%%%%%%%%%%%%%%%%%%%%%%%%
\section{R\'{e}nyi entropy based estimation theory and R\'{e}nyi entropy powers\label{sec2}}
%%%%%%%%%%%%%%%%%%%%%%%%%%%%%%%%%%%%%%%%%%%%%%%%%%%%%%%%%%%%%%%%%%%%%%%%%%%%%%%%%%%%%%%%%%%%%

In this section we introduce the concept of R\'{e}nyi's EP. With quantum metrology applications in mind, we discuss this in the framework of estimation theory. This will not only allow us to find new estimation inequalities, such as
the R\'{e}nyi-entropy-based De Bruijn identity, isoperimetric inequality or Stam inequality, but
it will also provide a convenient technical and conceptual frame for deriving a one-parameter
family of R\'{e}nyi-entropy-power-based quantum-mechanical uncertainty relations.

%%%%%%%%%%%%%%%%%%%%%%%%%%%%%%%%%%%%%%%%%%%%%%%%%%%%%%%%%%%%%%%%%%%%%%%%%%%%%%%%
\subsection{Fisher information --- Shannon's entropy approach\label{subsec2a}}
%%%%%%%%%%%%%%%%%%%%%%%%%%%%%%%%%%%%%%%%%%%%%%%%%%%%%%%%%%%%%%%%%%%%%%%%%%%%%%%%

First, we recall that the Fisher information matrix ${\mathbb{J}}(\mathcal{X})$ of a random vector $\{\mathcal{X}_i\}$
in ${\mathbb{R}}^D$ with the PDF $\mathcal{F}({\boldsymbol{x}})$ is defined as~\cite{frieden}
\begin{eqnarray}
{\mathbb{J}}(\mathcal{X}) \ = \ \mbox{cov}({\boldsymbol{V}}(\mathcal{X}))\, ,
\end{eqnarray}
where the covariance matrix is associated with the random zero-mean vector --- so-called {\em score vector}, as
%(i.e., log-derivative of the PDF)
%
\begin{eqnarray}
{\boldsymbol{V}}({\boldsymbol{x}}) \ = \  {\nabla \mathcal{F}({\boldsymbol{x}})}/{\mathcal{F}({\boldsymbol{x}})}\, .
\end{eqnarray}
A corresponding trace of  ${\mathbb{J}}(\mathcal{X})$, i.e.
\begin{eqnarray}
J(\mathcal{X}) \ = \ \mbox{Tr}({\mathbb{J}}(\mathcal{X})) \ = \ \mbox{var}({\boldsymbol{V}}(\mathcal{X})) \ = \
\mathbb{E}({\boldsymbol{V}}^2(\mathcal{X}))\, ,
\end{eqnarray}
is known as the Fisher information. Both the FI and FI matrix can be conveniently related to Shannon's differential entropy via De~Bruijn's identity~\cite{Rioul:11,Dembo:91}. \\[2mm]
{\em{De~Bruijn's identity}:} Let $\{\mathcal{X}_i\}$ be a random vector in ${\mathbb{R}}^D$ with the PDF $\mathcal{F}({\boldsymbol{x}})$
and let $\{\mathcal{Z}_i^{_G}\}$ be a Gaussian random vector (noise vector) with zero mean and unit-covariance matrix, independent of $\{\mathcal{X}_i\}$. Then
\begin{eqnarray}
\frac{d}{d\epsilon} {\mathcal{H}}(\mathcal{X} + \sqrt{\epsilon}\ \! \mathcal{Z}^{_G})|_{\epsilon = 0} \ = \ \frac{1}{2} J(\mathcal{X})\, ,
\end{eqnarray}
where
\begin{eqnarray}
{\mathcal{H}}(\mathcal{X})  \ = \  -\int_{
\mathbb{R}^{D}}\mathcal{F}({\boldsymbol{x}})\log \mathcal{F}
({\boldsymbol{x}})\ d{\boldsymbol{x}}\, ,
\end{eqnarray}
is Shannon's differential entropy (measured in {\em nats}). In the case when the independent additive noise  $\{\mathcal{Z}_i\}$ is non-Gaussian with zero mean and covariance matrix ${\boldsymbol{\Sigma}} = \mbox{cov}(\mathcal{Z})$ then the following generalization holds~\cite{Dembo:91}
\begin{eqnarray}
\frac{d}{d\epsilon} {\mathcal{H}}(\mathcal{X} + \sqrt{\epsilon}\ \! \mathcal{Z})|_{\epsilon = 0} \ = \ \frac{1}{2}\mbox{Tr}\left({\mathbb{J}}(\mathcal{X}) {\boldsymbol{\Sigma}}\right).
\label{DBI.5a}
\end{eqnarray}

%This can be naturally generalized to the case

The key point about De~Bruijn's identity is that it provides a very useful intuitive interpretation of FI, namely, FI quantifies the sensitivity of transmitted (Shannon type) information to an arbitrary independent additive noise. An important aspect that should be stressed in this context is that FI as quantifier of sensitivity depends only on the covariance of the noise vector and so it is independent of the shape of the noise distribution. This is because De~Bruijn's identity remains unchanged for both Gaussian and non-Gaussian additive noise with the same covariance matrix.

%%%%%%%%%%%%%%%%%%%%%%%%%%%%%%%%%%%%%%%%%%%%%%%%%%%%%%%%%%%%%%%%%%%%%%%%%%%%%%%%
\subsection{Fisher information --- R\'{e}nyi's entropy approach\label{subsec2b}}
%%%%%%%%%%%%%%%%%%%%%%%%%%%%%%%%%%%%%%%%%%%%%%%%%%%%%%%%%%%%%%%%%%%%%%%%%%%%%%%%

We now extend the notion of the FI matrix to the R\'{e}nyi entropy setting. A natural way to do it is via an extension of De~Bruijn's identity to R\'{e}nyi entropies.
%
%prove that the shape of the noise distribution can be further quantified when ****
%
%
%by the R\'{e}nyi-entropy-based FI matrix.
In particular, the following statement holds: \\[2mm]
{\em{Generalized De~Bruijn's identity}:} Let $\{\mathcal{X}_i\}$ be a random vector in ${\mathbb{R}}^D$ with the PDF $\mathcal{F}({\boldsymbol{x}})$
and let $\{\mathcal{Z}_i\}$ be an independent (generally non-Gaussian) noise vector with the zero mean and covariance matrix ${\boldsymbol{\Sigma}} = \mbox{cov}(\mathcal{Z})$, then
\begin{eqnarray}
\frac{d}{d\epsilon} {\mathcal{I}}_q(\mathcal{X} + \sqrt{\epsilon}\ \! \mathcal{Z})|_{\epsilon = 0} \ = \  \frac{1}{2q}\mbox{Tr}\left({\mathbb{J}}_q(\mathcal{X}) {\boldsymbol{\Sigma}}\right)\, ,
\label{DeB.2a}
\end{eqnarray}
where
\begin{eqnarray}
{\mathcal{I}}_q  \ = \  \frac{1}{1-q}\log\int_{\mathbb{R}^{D}}{\mathcal{F}}^q({\boldsymbol{x}}) d{\boldsymbol{x}}\, ,
\end{eqnarray}
is  {\em R\'{e}nyi's differential entropy} (measured in {\em nats}) with  ${\mathcal{I}}_1 = {\mathcal{H}}$.
The ensuing FI matrix of order $q$ has the explicit form
\begin{eqnarray}
{\mathbb{J}}_q(\mathcal{X}) \ = \ \mbox{cov}_q({\boldsymbol{V}}_q(\mathcal{X})) \, ,
\end{eqnarray}
with the score vector
\begin{eqnarray}
{\boldsymbol{V}}_q({\boldsymbol{x}}) \ = \ {\nabla \rho_q({\boldsymbol{x}})}/{\rho_q({\boldsymbol{x}})}
 \ = \ q \nabla {\mathcal{F}}({\boldsymbol{x}})/{\mathcal{F}}({\boldsymbol{x}}) \ = \ q {\boldsymbol{V}}({\boldsymbol{x}})   \, .
\end{eqnarray}
Here $\rho_q = {\mathcal{F}}^q/\int_{\mathbb{R}^{D}}{\mathcal{F}}^q d{\boldsymbol{x}}$ is the so-called {\em escort distribution}~\cite{beck:93}.
The ``$\mbox{cov}_q$'' denotes the covariance matrix computed with respect to $\rho_q$.
Proofs of both the conventional (i.e., Shannon entropy based) and generalized (i.e., R\'{e}nyi entropy based) De~Bruijn's identity are provided in Appendix~A.
There we also discuss some further useful generalizations of De~Bruijn's identity. Finally, as in the Shannon case we define the FI of order $q$ --- denoted as $J_q(\mathcal{X})$, as
\begin{eqnarray}
\mbox{Tr}\left({\mathbb{J}}_q(\mathcal{X}) \right) \ \equiv \  J_q(\mathcal{X})\, .
\label{DeB.2ab}
\end{eqnarray}

%%%%%%%%%%%%%%%%%%%%%%%%%%%%%%%%%%%%%%%%%%%%%%%%%%%%%%%%%%%%%%%%%%%%%%%%%%%%%%%%%%%%%%%%%%%%%%%%
\subsection{R\'{e}nyi's entropy power and generalized isoperimetric inequality \label{subsec2c}}
%%%%%%%%%%%%%%%%%%%%%%%%%%%%%%%%%%%%%%%%%%%%%%%%%%%%%%%%%%%%%%%%%%%%%%%%%%%%%%%%%%%%%%%%%%%%%%%%%

Similarly as in conventional estimation theory, one can expect that there should exist a close connection between the FI matrix ${\mathbb{J}}_q(\mathcal{X})$ and the corresponding R\'{e}nyi entropy power $N_p(\mathcal{X})$. In Shannon's information theory this
such a connection is phrased in terms of isoperimetric inequality~\cite{Dembo:91}. Here we prove that a similar relationship works also in R\'{e}nyi's information theory.

Let us start by introducing the concept of R\'{e}nyi's entropy power. This is defined as the
%~\cite{JDJ,Gardner02}:
solution of the equation~\cite{JD:16,JDJ}
\begin{eqnarray}
{\mathcal{I}_{p}} \left( {\mathcal{X}} \right)
\ = \ \mathcal{I}_{p}\left(\sqrt{N_p(\mathcal{X})}\cdot
{\mathcal{Z}}^{_G}\right)\, ,
\label{3.1.0k}
\end{eqnarray}
where $\{{\mathcal{Z}}_{i}^{_G}\}$ represents a Gaussian random vector with zero mean
and unit covariance matrix. So, $N_p(\mathcal{X})$ denotes the variance of a would be Gaussian distribution that
has the same R\'{e}nyi information content as the random vector $\{\mathcal{X}_i\}$ described by the PDF ${\mathcal{F}}({\boldsymbol{x}})$.
%Alternatively, one may view the EP introduced in (\ref{3.1.0k}) as the effective support set size for a random vector.
\textit{}Expression (\ref{3.1.0k}) was studied in~\cite{JD:16,JDJ,Gardner02} where it was shown
that the only  class of solutions of (\ref{3.1.0k}) is
\begin{eqnarray}
N_p(\mathcal{X})
% &=& \frac{1}{2\pi} p^{-p'/p} |\!|
% {\mathcal{F}}|\!|_p^{-2p'/D} \nonumber \\
\ = \ \frac{1}{2\pi} p^{-p'/p}
\exp\left(\frac{2}{D} \ \!{\mathcal{I}}_p({\mathcal{X}})\right),
\label{3.1.0e}
\end{eqnarray}
with $1/p + 1/p'= 1$ and $p\in {\mathbb{R}}^{+}$.
%(i.e., $p'$ and $p$ are H\"{o}lder conjugates).
In addition, when $p \rightarrow 1_+$ one has
$N_p(\mathcal{X}) \rightarrow N(\mathcal{X})$, where $N(\mathcal{X})$ is the conventional Shannon entropy power~\cite{Shannon48}.
In this latter case one can use the {\em asymptotic equipartition
property}~\cite{Cover:06,JK-PRL:19} to identify $N(\mathcal{X})$  with ``typical size'' of a state set, which in the present context is the effective support set size for a random vector. This, in turn, is equivalent to Einstein's entropic principle~\cite{Einstein:10}.
In passing, it should be noted that the form of the R\'{e}nyi EP expressed in (\ref{3.1.0e}) is not universally accepted version.  In a number of works it is defined merely as an exponent of RE, see, e.g.~\cite{DePalma,Ram}.
Our motivation for the form (\ref{3.1.0e}) is twofold: first, it has a clear interpretation in terms of variances of Gaussian distributions and second, it leads to simpler formulas, cf. e.g., Eq.~(\ref{REPURaa}).
\\[2mm]
{\em{Generalized isoperimetric inequality}:}
Let $\{\mathcal{X}_i\}$ be a random vector in ${\mathbb{R}}^D$ with the PDF $\mathcal{F}({\boldsymbol{x}})$. Then
\begin{eqnarray}
\frac{1}{D} N_q(\mathcal{X}) J_q(\mathcal{X})\ \geq \ N_q(\mathcal{X}) [\det({\mathbb{J}}_q(\mathcal{X}))]^{1/D} \ \geq \ 1\, ,
\label{2.12ac}
\end{eqnarray}
where the R\'{e}nyi parameter $q \geq 1 $. We relegate the proof of the generalized isoperimetric inequality to Appendix~B.

It is also worth noting that the relation (\ref{2.12ac}) implies another important inequality.
By using the fact that the Shannon entropy is maximized (among all PDF's with identical covariance matrix $\boldsymbol{\Sigma}$)
by the Gaussian distribution we have $N_1(\mathcal{X}) \leq
\det(\boldsymbol{\Sigma})^{1/D}$ (see, e.g.~\cite{Stam:59}). If we further employ that ${\mathcal{I}_{q}}$ is a monotonously decreasing function of $q$, see, e.g.~\cite{Jizba2004,Renyi1970},  we can write (recall that $q \geq 1$)
\begin{eqnarray}
 \frac{q^{1/(q-1)}}{e}\ \! N_q \ \leq \ N_1  \ = \ \frac{\exp(\frac{2}{D} {\mathcal{I}_{1}})}{2\pi e} \ \leq \
\det(\boldsymbol{\Sigma})^{1/D} .
\end{eqnarray}
The isoperimetric inequality (\ref{2.12ac}) then implies
\begin{eqnarray}
\det(\boldsymbol{\Sigma}(\mathcal{X})) \ \geq \ \frac{\left(q^{1/(q-1)}\right)^D}{e^D\det({\mathbb{J}}_q(\mathcal{X}))}
\ \geq\  \frac{1}{e^D\det({\mathbb{J}}_q(\mathcal{X}))}\, .
\label{II.14.aa}
\end{eqnarray}
We can further use the inequality
%of arithmetic and geometric means which implies that
%
\begin{eqnarray}
\frac{1}{D} \mbox{Tr}({\mathbb{A}}) \ \geq \ [\det({\mathbb{A}})]^{1/D}\, ,
\end{eqnarray}
(valid for any positive semi-definite $D\times D$ matrix ${\mathbb{A}}$) to write
\begin{eqnarray}
\sigma^{2}(\mathcal{X})  \ = \ \frac{1}{D} \mbox{Tr}(\boldsymbol{\Sigma}(\mathcal{X})) \ = \ \frac{1}{D} \sum_{i=1}^D \mbox{Var}(\mathcal{X}_i)
\ \geq \ \frac{D q^{1/(q-1)}}{e J_q(\mathcal{X})} \ \geq \
\frac{D}{e J_q(\mathcal{X})}\, ,
\label{II.16.aa}
\end{eqnarray}
where $\sigma^{2}$ is an average variance per component.

Relations (\ref{II.14.aa})-(\ref{II.16.aa}) represent the $q$-generalizations of the celebrated Cram\'{e}r--Rao information
inequality. In the limit of $q \rightarrow 1$ we recover the standard
Cram\'{e}r--Rao inequality which is widely used in statistical inference theory~\cite{Cramer:46,frieden}.
A final logical step needed to complete the proof of REPURs is represented by the so called generalized Stam inequality. To this end we first define the concept
of {\em conjugate random variables}. We say that random vectors $\{\mathcal{X}_i\}$ and $\{\mathcal{Y}_i\}$ in $\mathbb{R}^D$
are conjugate if their respective PDF's $\mathcal{F}({\boldsymbol{x}})$ and  $\mathcal{G}({\boldsymbol{y}})$ can be written as
\begin{eqnarray}
\mathcal{F}({\boldsymbol{x}}) \ = \ |\varphi_{_{\mathcal{F}}}({\boldsymbol{x}})|^2/|\!|\varphi_{_{\mathcal{F}}}|\!|_2^2
% \ = \  |\hat{\psi}_{_{\mathcal{G}}}({\boldsymbol{x}})|^2/|\!|\hat{\psi}_{_{\mathcal{G}}}|\!|_2^2
 \, , \;\;\;\;\;\;\;
\mathcal{G}({\boldsymbol{y}}) \ = \ |\varphi_{_{\mathcal{G}}}({\boldsymbol{y}})|^2/|\!|\varphi_{_{\mathcal{G}}}|\!|_2^2
% \ = \  |\hat{\psi}_{_{\mathcal{F}}}({\boldsymbol{y}})|^2/|\!|\hat{\psi}_{_{\mathcal{F}}}|\!|_2^2
 \, ,
\label{19.ac}
\end{eqnarray}
where the (generally complex) probability amplitudes $\varphi_{_{\mathcal{F}}}({\boldsymbol{x}}) \in L_2(\mathbb{R}^D)$ and $\varphi_{_{\mathcal{G}}}({\boldsymbol{y}}) \in L_2(\mathbb{R}^D)$ are mutual Fourier images, i.e.,
\begin{eqnarray}
\varphi_{_{\mathcal{F}}}({\boldsymbol{x}}) \ = \  \hat{\varphi}_{_{\mathcal{G}}}({\boldsymbol{x}}) \ = \ \int_{\mathbb{R}^{D}}e^{2\pi
i{{\boldsymbol{x}} }.{{\boldsymbol{y}}}}\
\varphi_{_{\mathcal{G}}}({{\boldsymbol{y}}})\ d{{\boldsymbol{y}}}\, ,
\end{eqnarray}
and analogously for $\varphi_{_{\mathcal{G}}}({\boldsymbol{y}}) = \hat{\varphi}_{_{\mathcal{F}}}({\boldsymbol{y}})$. With this we can state the generalized Stam inequality. \\[2mm]
{\em{Generalized Stam inequality}:} Let $\{\mathcal{X}_i\}$ and $\{\mathcal{Y}_i\}$ be conjugate random vectors in ${\mathbb{R}}^D$. Then
\begin{eqnarray}
16 \pi^2 N_q({\mathcal{Y}}) \  \geq \  [\det({\mathbb{J}}_r(\mathcal{X}))]^{1/D}\, ,
\label{stam.ineq.a}
\end{eqnarray}
is valid for any $r \in [1,\infty)$ and $q \in [1/2, 1]$ that are connected via the relation $1/r + 1/q =2$ (i.e., $r/2$ and $q/2$  are H\"{o}lder conjugates). A  proof of the generalized Stam inequality is provided in Appendix~C.

Combining the isoperimetric inequality (\ref{2.12ac}) together with the generalized Stam inequality (\ref{stam.ineq.a}) we obtain a one-parameter class of REP-based inequalities
\begin{eqnarray}
%\mbox{\hspace{-3mm}}N_{1+t}({\mathcal{F}}^{(2)})N_{1+r}({\mathcal{F}}^{(1)})  \equiv
N_{p/2}({\mathcal{X}})N_{q/2}({\mathcal{Y}}) &=& N_{p/2}({\mathcal{X}})\frac{[\det({\mathbb{J}}_{p/2}(\mathcal{X}))]^{1/D}}{[\det({\mathbb{J}}_{p/2}(\mathcal{X}))]^{1/D}} N_{q/2}({\mathcal{Y}})  \nonumber \\[2mm] &\geq&    \frac{N_{q/2}({\mathcal{Y}})}{[\det({\mathbb{J}}_{p/2}(\mathcal{X}))]^{1/D}} \ \geq  \  \frac{1}{16\pi^2}\, ,
\label{REPURaa}
\end{eqnarray}
where $p$ and $q$ form now a H\"{o}lder double. By symmetry the role of $q$ and $p$ can be reversed. In~Refs.~\cite{JD:16,JDJ} we presented an alternative derivation of inequalities (\ref{REPURaa}) that was based on the Beckner--Babenko theorem. There it was also proved that the inequality saturates if and only if the distributions involved are Gaussian. The only exception to this rule is for the asymptotic values $p= 1$ and $q = \infty$ (or vice versa) where the
saturation happens whenever the peak of $\mathcal{F}({\boldsymbol{x}})$ and tail of $\mathcal{G}({\boldsymbol{y}})$ (or vice versa) are Gaussian.

The passage to quantum mechanics is quite straightforward.  First, we realize that in QM the Fourier conjugate wave functions are related
via two reciprocal relations
\begin{eqnarray}
&&\psi_{_{\mathcal{F}}}({\bf{x}})  \ = \ \int_{\mathbb{R}^D} e^{i {\bf y}\cdot {\bf x}/\hbar} \  \! {\psi}_{_{\mathcal{G}}}({\bf{y}})\ \!  \frac{d{\bf y}}{(2\pi \hbar)^{D/2}}\, ,\nonumber \\[2mm]
&&{\psi}_{_{\mathcal{G}}}({\bf{y}})  \ = \ \int_{\mathbb{R}^D} e^{-i {\bf y}\cdot {\bf x}/\hbar} \  \! {\psi}_{_{\mathcal{F}}}({\bf{x}})\ \!  \frac{d{\bf x}}{(2\pi \hbar)^{D/2}}\, . \label{V.1.a}
\end{eqnarray}
The Plancherel (or Riesz--Fischer) equality  implies that when $|\!|{\psi}_{_{\mathcal{F}}}|\!|_2 = 1$ then also automatically $|\!|
{\psi}_{_{\mathcal{G}}}|\!|_{2} = 1$ (and vice versa).  So, the connection between amplitudes  $\varphi_{_{\mathcal{F}}}$ and $\varphi_{_{\mathcal{G}}}$ from (\ref{19.ac}) and amplitudes $\psi_{_{\mathcal{F}}}$ and ${\psi}_{_{\mathcal{G}}}$ from (\ref{V.1.a})  is
\begin{eqnarray}
&&\varphi_{_{\mathcal{F}}}({\bf{x}}) \ = \ (2\pi \hbar)^{D/4}\psi_{_{\mathcal{F}}}(\sqrt{2\pi\hbar}\ \! {\bf{x}})\, , \nonumber \\[2mm]
&&\varphi_{_{\mathcal{G}}}({\bf{y}}) \ = \ (2\pi \hbar)^{D/4}{\psi_{_{\mathcal{G}}}}(\sqrt{2\pi\hbar}\ \! {\bf{y}}) \, .
\label{6.2.a}
\end{eqnarray}
The factor $(2\pi \hbar)^{D/4}$ ensures that also $\varphi_{_{\mathcal{F}}}$ and $\varphi_{_{\mathcal{G}}}$ functions are normalized (in sense of $|\!|\ldots|\!|_2$) to unity, however, due to Eq.~(\ref{19.ac}) it might be easily omitted. The corresponding R\'{e}nyi EPs change according to
\begin{eqnarray}
&&N_{p/2}({\mathcal{X}}) \ \equiv \ N_{p/2}({\mathcal{F}}) \;\; \mapsto \;\; N_{p/2}(|\psi_{_{\mathcal{F}}}|^2) \ = \   2\pi \hbar N_{p/2}({\mathcal{F}})\, , \nonumber \\[2mm]
&& N_{q/2}({\mathcal{Y}}) \ \equiv \ N_{q/2}({\mathcal{G}}) \;\; \mapsto \;\; N_{q/2}(|\psi_{_{\mathcal{G}}}|^2) \ = \   2\pi \hbar N_{q/2}({\mathcal{G}})\, ,
\end{eqnarray}
and hence REP-based inequalities (\ref{REPURaa}) acquire in the QM setting a simple form
\begin{eqnarray}
N_{p/2}(|\psi_{_{\mathcal{F}}}|^2)N_{q/2}(|\psi_{_{\mathcal{G}}}|^2) \ \geq \ \frac{\hbar^2}{4}\, .
\label{26bc}
\end{eqnarray}
This represents an infinite tower of mutually distinct (generally irreducible) REPURs~\cite{JD:16}.

At this point, some comments are in order. First, historically the most popular quantifier of quantum uncertainty has been {\em variance} because it is conceptually simple and relatively easily extractable from experimental data.
The variance determines the measure of uncertainty in terms of the fluctuation (or spread) around
the mean value which, while useful for many distributions, does not provide a sensible
measure of uncertainty in a number of important situations including multimodal~\cite{Birulla:07,JDJ,JD:16} and
heavy-tailed distributions~\cite{JDJ,JD:16,Maasen:88}. To deal with this, a multitude of alternative (non-variance based) measures of uncertainty in quantum mechanics (QM) have emerged. Among these, a particularly prominent role is played by information entropies such as the Shannon entropy~\cite{Bengtsson}, R\'{e}nyi entropy~\cite{Bengtsson,JDJ}, Tsallis entropy~\cite{Wilk}, associated differential entropies
and their quantum-information generalizations~\cite{Birula,JD:16,JDJ}. REPURs (\ref{26bc}) fit into this framework of entropic QM URs.
In connection with (\ref{26bc}) one might observe that the conventional URs based on variances --- so-called Robertson--Schr\"{o}dinger URs~\cite{Schroedinger1,Robertson1929}) and Shannon differential entropy based URs (e.g., Hirschman or Bia{\l}ynicki-Birula  URs~\cite{bourret:58,Birula}) naturally appear as special cases in this hierarchy.
Second, the ITEs  enter  quantum  information theory  typically  in  three  distinct  ways:
a)  as  a  measure of the quantum information content (e.g., how many qubits are needed to encode  the  message  without  loss  of  information), b) as a measure of the classical information content (e.g., amount of information in bits that can be recovered from the quantum system) and c) to quantify the entanglement of pure and mixed bipartite quantum states.
Logarithms in base 2 are used because in quantum information, one quantifies entropy in bits and qubits (rather than nats).
This in turn also modifies R\'{e}nyi's EP as

\begin{eqnarray}
\frac{1}{2\pi} p^{-p'/p}
e^{\left(\frac{2}{D} \ \!\cdots \right)}\;\; \mapsto \;\; \frac{1}{2\pi} p^{-p'/p} \ \!
2^{\left(\frac{2}{D} \ \!\cdots \right)}\, .
\label{27.e}
\end{eqnarray}
In the following we will employ this QM practice.

%In an earlier paper~\cite{JD:16}, we introduced an infinite tower of mutually distinct (generally irreducible)
%R\'{e}nyi entropy-power-based URs (REPURs).

%%%%%%%%%%%%%%%%%%%%%%%%%%%%%%%%%%%%%%%%%%%%%%%%%%%%%%%%
\section{Information distribution\label{sec3}}
%%%%%%%%%%%%%%%%%%%%%%%%%%%%%%%%%%%%%%%%%%%%%%%%%%%%%%%%

To put more flesh on the concept of R\'{e}nyi's EP, we devote the rest of this paper to the
development of the methodology and application of R\'{e}nyi EPs in extracting quantum states
from incomplete  data. This is of particular interest in various quantum metrology protocols. To this end we first start with the notion of the
information distribution. \\[1mm]

Let $\mathcal{F}({\boldsymbol{x}})$ be the PDF for the random variable ${\mathcal{X}}$.
We define the {\em information random variable}
$i_{{\mathcal{X}}}({\mathcal{X}})$ so that $i_{{\mathcal{X}}}({\boldsymbol{x}}) = \log_2 1/\mathcal{F}({\boldsymbol{x}})$. In other words
$i_{{\mathcal{X}}}({\boldsymbol{x}})$ represents the information in ${\boldsymbol{x}}$ with
respect to $\mathcal{F}({\boldsymbol{x}})$.
In this connection it is expedient to introduce the cumulative distribution function for $i_{{\mathcal{X}}}({\mathcal{X}})$ as
\begin{eqnarray}
\!\wp(y)  =  \int_{-\infty}^y d\wp(i_{{\mathcal{X}}})
  =  \int_{\mathbb{R}^D} \mathcal{F}({\boldsymbol{x}}) \theta(\log_2 \mathcal{F}({\boldsymbol{x}}) + y) d{\boldsymbol{x}}\, .
 \label{SEc5c.7aa}
\end{eqnarray}
The function $\wp(y)$ thus represents the probability that the random variable
$i_{{\mathcal{X}}}({\mathcal{X}})$
is less or equal than $y$. We have denoted the corresponding  probability measure as  $ d\wp(i_{{\mathcal{X}}})$.
Taking the  Laplace transform of both sides of (\ref{SEc5c.7aa}), we get
\begin{eqnarray}
\!\!\mathcal{L}\{\wp\}(s)   \ =  \   \int_{\mathbb{R}^D}\!\!\! \mathcal{F}({\boldsymbol{x}}) \ \! \frac{e^{s\log_2 \mathcal{F}({\boldsymbol{x}}) }}{s}
\ \! d{\boldsymbol{x}}  \ = \  \frac{{\mathbb{E}}\left[e^{s\log_2 \mathcal{F} }  \right]}{s} \, ,
\end{eqnarray}
where $\mathbb{E}\left[\cdots \right]$ denotes the mean value with respect to $\mathcal{F}$.
By assuming that $\wp(x)$ is smooth then the PDF associated with $i_{{\mathcal{X}}}({\mathcal{X}})$ --- the so-called {\em information PDF} -- is
\begin{eqnarray}
g(y) \ = \ \frac{d\wp(y)}{dy} \ = \  \mathcal{L}^{-1}\left\{ {\mathbb{E}}\left[e^{s\log_2 \mathcal{F}} \right]  \right\}\!(y)\, .
\label{9aa}
\end{eqnarray}
Setting $s = (p-1)\log 2$ we have
\begin{eqnarray}
\mathcal{L}\{g\}(s = (p-1)\log 2) \ = \  \mathbb{E}\left[ 2^{(1-p)i_{{\mathcal{X}}}} \right] .
 \label{SEc5c.8aa}
\end{eqnarray}
The mean here is taken with respect to the PDF $g$. Eq.~(\ref{SEc5c.8aa}) can also be written explicitly as
\begin{eqnarray}
\int_{\mathbb{R}^D} d{{\boldsymbol{x}}} \ \! \mathcal{F}^{p}({\boldsymbol{x}}) \ = \ \int_{\mathbb{R}} g(y) 2^{(1-p)y}dy\, .
\label{II.11aa}
\end{eqnarray}
Note that when $\mathcal{F}^{p}$ is integrable for $p\in [1,2]$  then (\ref{II.11aa})
ensures that the moment-generating function for $g(x)$ PDF exists. So in particular, the moment-generating
function exists when $\mathcal{F}({\boldsymbol{x}})$ represents L\'{e}vy $\alpha$-stable distributions, including the
heavy-tailed stable distributions (i.e, PDFs with the L\'{e}vy stability parameter $\alpha \in(0,2]$). The same holds
for $\hat{\mathcal{F}}$ and $p'\in [2,\infty)$  due to the Beckner--Babenko theorem~\cite{Beckner1975,Babenko1962,JD:16}.

%%%%%%%%%%%%%%%%%%%%%%%%%%%%%%%%%%%%%%%%%%%%%%%%%%%%
\section{Reconstruction theorem\label{sec4}}
%%%%%%%%%%%%%%%%%%%%%%%%%%%%%%%%%%%%%%%%%%%%%%%%%%%%%

Since $\mathcal{L}\{g\}(s)$ is the {\em  moment-generating function} of the random variable $i_{{\mathcal{X}}}({\mathcal{X}})$
one can generate all moments of the PDF $g(x)$ (if they exist) by taking the derivatives of $\mathcal{L}\{g\}$ with respect to $s$.
From a conceptual standpoint, it is often more useful to work with cumulants rather than moments.  Using the fact that the
{\em cumulant generating function}  is simply the (natural) logarithm of the moment-generating function,
we  see from (\ref{II.11aa}) that the differential RE is a reparametrized version of
the cumulant generating function of the information random variable $i_{{\mathcal{X}}}({\mathcal{X}})$.
In fact, from (\ref{SEc5c.8aa}) we have
\begin{equation}
{\mathcal{I}}_p(\mathcal{X}) \ = \ \frac{1}{(1-p)}\log_{2} \mathbb{E}\left[ 2^{(1-p)i_{{\mathcal{X}}}} \right]  \, . \label{SEc5c.1}
\end{equation}
To understand the meaning of  REPURs we begin with
the cumulant expansion (\ref{SEc5c.1}), i.e.
\begin{eqnarray}
p\mathcal{I}_{1-p}({\mathcal{X}}) \ = \  \log_2 e \sum_{n=1}^{\infty}
 \frac{\kappa_n({\mathcal{X}})}{n!} \left(\frac{p}{\log_2 e}\right)^n\! ,
\label{SEc5c.1a}
\end{eqnarray}
where $\kappa_n({\mathcal{X}}) \equiv \kappa_n(i_{{\mathcal{X}}})$ denotes the $n$-th cumulant of the information random variable $i_{{\mathcal{X}}}({\mathcal{X}})$ (in units of {\em bits}$^n$). We note that
\begin{eqnarray}
&&\kappa_1({\mathcal{X}}) \ = \  \mathbb{E}\left[i_{{\mathcal{X}}}({\mathcal{X}})\right] \ = \ \mathcal{H}({\mathcal{X}})\, ,\nonumber \\[1mm]
&&\kappa_2({\mathcal{X}}) \ = \ \mathbb{E}\left[i_{{\mathcal{X}}}({\mathcal{X}})^2\right]- (\mathbb{E}\left[i_{{\mathcal{X}}}({\mathcal{X}})\right])^2\, ,
\end{eqnarray}
i.e., they represent the Shannon entropy and {\em varentropy}, respectively. By employing the identity
\begin{eqnarray}
\mbox{\hspace{-0.9cm}}\mathcal{I}_{1-p}({\mathcal{X}})
%\ &=& \ \mathcal{I}_{1-p}(\sqrt{N_{1-p}(\mathcal{X})}\cdot
%{\mathcal{Z}}_{G})\nonumber \\
%\ = \ \mathcal{I}_{1-p}({\mathcal{Z}}_{G}) \ + \ \frac{D}{2} \log_2 N_{1-p}(\mathcal{X})
&=& \  \frac{D}{2} \log_2 \left[
2\pi(1-p)^{-1/p} N_{1-p}(\mathcal{X})\right]
\! ,
\label{SEc5c.2a}
\end{eqnarray}
we can rewrite (\ref{SEc5c.1a}) in the form
\begin{eqnarray}
\log_2 \left[ N_{1-p}(\mathcal{X})\right]
\ =  \ \log_2 \left[\frac{(1-p)^{1/p}}{2\pi} \right]
 \ + \   \frac{2}{D} \sum_{n=1}^{\infty} \frac{\kappa_n({\mathcal{X}})}{n!} \left(\frac{p}{\log_2 e}\right)^{n-1}\!\! .
\label{SEc5c.3a}
\end{eqnarray}
From (\ref{SEc5c.3a}) one can see that
\begin{eqnarray}
\mbox{\hspace{-8mm}}\kappa_n({\mathcal{X}}) \ &=& \ \left.  \frac{nD}{2}(\log_2 e)^{n-1} \frac{d^{n-1}
\log_2 \left[ N_{1-p}(\mathcal{X})\right]}{dp^{n-1}}\right|_{p=0}
\ +  \ \frac{D}{2}(\log_2 e)^n \left[(n-1)! + \delta_{1n}\log 2\pi\right]\, ,
\label{SEc5c.4aa}
\end{eqnarray}
which, in terms of the Gr\"{u}nwald-Letnikov derivative formula (GLDF)~\cite{Samko}, allows us to write
\begin{eqnarray}
\kappa_n({\mathcal{X}}) &=&  \lim_{\varDelta \rightarrow 0}\frac{nD}{2}
\frac{(\log_2 e)^{n} }{\varDelta^{n-1}}\ \! \sum_{k=0}^{n-1} (-1)^k \binom{n-1}{k} \log
\left[ N_{1+k\varDelta }(\mathcal{X})\right]\nonumber \\[2mm]
&& + \ \frac{D}{2}(\log_2 e)^n \left[(n-1)! \ + \ \delta_{1n}\log 2\pi\right]\, .
\label{SEc5c.5aa}
\end{eqnarray}
So, in order to determine the first $m$ cumulants of $i_{{\mathcal{X}}}({\mathcal{X}})$ we need to know all
$N_{1}, N_{1+\varDelta}, \ldots, N_{1+(m-1)\varDelta}$ entropy powers. In practice $\varDelta$ corresponds
to a characteristic resolution scale for the entropy index which will be chosen appropriately for the task at hand,
but is typically of the order $10^{-2}$. Note that the last term in (\ref{SEc5c.4aa}) and (\ref{SEc5c.5aa}) can be also written
\begin{eqnarray}
\frac{D}{2}(\log_2 e)^n \left[(n-1)! + \delta_{1n}\log 2\pi\right] \
\ = \ \kappa_n({\mathcal{Z}}_{G}^{\ide}) \ \equiv \ \kappa_n(i_{{\mathcal{Y}}})\, ,
\end{eqnarray}
with ${{\mathcal{Y}}}$ being the random variable distributed with respect
to the {\em Gaussian} distribution ${\mathcal{Z}}_{G}^{\ide}$ with the {\em unit} covariance matrix.
%Analogously we can formulate (\ref{SEc5c.5aaa}) in terms of the GLDF.

When all the cumulants exist then the problem of recovering the underlying PDF for $i_{{\mathcal{X}}}({\mathcal{X}})$
is equivalent to the {\em  Stieltjes} moment problem~\cite{Reed1975}. Using this connection, there are a number of ways to proceed; the PDF in question can be
reconstructed e.g.,  in terms of sums involving orthogonal polynomials (e.g., the Gram--Charlier A series or the Edgeworth series~\cite{Wallaca:58}), the inverse Mellin transform~\cite{Zolotarev}
or via various maximum entropy techniques~\cite{Frontini:98}.
Pertaining to this, the theorem of Marcinkiewicz~\cite{Lukacz} implies that
there are no PDFs for which $\kappa_m = \kappa_{m+1} = \ldots = 0$ for $m \geq 3$. In other words,
the cumulant generating function cannot be a finite-order polynomial of degree greater than $2$.
The important exceptions, and indeed the only exceptions to Marcinkiewicz's theorem, are the {\em Gaussian} PDFs which can have
the first two cumulants nontrivial and $\kappa_3 = \kappa_4 = \ldots = 0$.
Thus, apart from the special case of Gaussian PDFs where only $N_{1}$ and $N_{1+\varDelta}$ are needed,
one needs to work with as many entropy powers $N_{1+k\varDelta}, k\in {\mathbb{N}}$ (or ensuing REPURs) as possible
to receive as much information as possible about the structure of the underlying PDF. In theory, the whole infinite
tower of REPURs would be required to uniquely specify a system's information PDF.
Note that for {\em Gaussian} information PDFs  one needs only $N_{1}$ and $N_{1+\varDelta}$ to
reconstruct the PDF uniquely. From (\ref{SEc5c.3a}) and (\ref{SEc5c.5aa}) we see that knowledge of $N_1$ corresponds to
$\kappa_1({\mathcal{X}}) =  \mathcal{H}({\mathcal{X}})$ while $N_{1+\varDelta}$ further determines $\kappa_2$, i.e. the varentropy. Since $N_1$ is involved (via (\ref{SEc5c.5aa}))
in the determination of all cumulants, it is the most important entropy power in the tower.
So, the entropy powers of a given process have an equivalent meaning to the PDF: they
describe the morphology of uncertainty of the observed phenomenon.

We should stress, that the focus of the reconstruction theorem we present
is on cumulants $\kappa_n$ which can be directly used for a shape estimation of $g(x)$ but not $\mathcal{F}({\boldsymbol{x}})$.
However, by knowing $g(y)$ we have a complete ``information scan'' of  $\mathcal{F}({\boldsymbol{x}})$.
Such an information scan is, however, not unique, indeed two PDFs that are rearrangements of each other -- i.e.,
{\em equimeasurable} PDFs, have identical $\wp(y)$ and $g(y)$.
Even though equimeasurable PDFs cannot be distinguished via their entropy powers, they can be, as a rule, distinguished via
their respective momentum-space PDFs and associated entropy powers. So the information scan has a tomographic flavor to it. From the multi-peak
structure of $g(y)$ one can determine the {\em number} and {\em height} of the stationary points.  These are invariant characteristics of a given
family of equimeasurable PDFs. This will be further illustrated in Sec.~VI.\\
%
%%%%%%%%%%%%%%%%%%%%%%%%%%%%%%%%%%%%%%%%%%%%%%%%%%%%%%%%
\section{Information scan of quantum-state PDF\label{sec5}}
%%%%%%%%%%%%%%%%%%%%%%%%%%%%%%%%%%%%%%%%%%%%%%%%%%%%%%%%

With knowledge of the entropy powers, the question now is how we can reconstruct the information distribution $g(x)$.
The inner workings of this will now be explicitly illustrated with the (generalized) Gram-Charlier~A expansion. However, other
-- often more efficient methods -- are also available~\cite{Wallaca:58}.
Let $\kappa_n$ be cumulants obtained from entropy powers and let $G({x})$ be some reference PDF whose cumulants are $\gamma_k$. The information PDF $g({x})$ can be then written as~\cite{Wallaca:58}
\begin{eqnarray}
\mbox{\hspace{-0.5cm}} g({x}) \ = \ \exp\left[\sum_{k=1}^{\infty}(\kappa_k - \gamma_k) (-1)^k \frac{ (d^k/dx^k)}{k!}   \right] G({x})\, .
\label{12abc}
\end{eqnarray}
With the hindsight  we choose the reference PDF $G({x})$ to be a shifted gamma PDF, i.e.
\begin{eqnarray}
\mbox{\hspace{-0.5cm}}G({x}) \ &\equiv&\ {\mathcal{G}}({x}|a,\alpha,\beta)  \ = \ \frac{e^{-(x-a)/\beta}(x-a)^{\alpha -1}}{{\beta^{\alpha} \Gamma[\alpha]}}\, ,
\end{eqnarray}
with $a < x < \infty, \; \beta > 0, \; \alpha > 0$.
In doing so, we have implicitly assumed that the $\mathcal{F}(y)$ PDF is in the first approximation equimeasurable with the Gaussian PDF.
 To reach a corresponding matching we should choose  $a =  \log_2 (2\pi \sigma^2)/2$, $\alpha = 1/2$
 and $\beta = \log_2 e$. Using the fact that~\cite{cc}
\begin{eqnarray}
{(\beta)^{k+1/2}} \ \! \frac{d^k\ \! {\mathcal{G}}({x}|a,1/2,\beta)}{{k!} \ \! dx^k} \
= \  \left(\frac{x-a}{\beta}\right)^{-k}L_k^{(-1/2-k)}\left(\frac{x-a}{\beta}\right) {\mathcal{G}}({x}|a,1/2,\beta)\, ,
\end{eqnarray}
(where $L_k^{\delta} $ is an associated Laguerre polynomial of order $k$ with parameter $\delta$) and given that
$\kappa_1 = \gamma_1 =  \alpha\beta + a = \log_2(2\pi \sigma^2 e)/2$,  and  $\gamma_k = \Gamma(k)\alpha \beta^k =  (\log_2 e)^k/2$ for $k>1$ we can write (\ref{12abc}) as
\begin{eqnarray}
\mbox{\hspace{-5mm}}g(x)   &=& {\mathcal{G}}({x}|a,1/2,\beta) \left[1  \ + \  \frac{(\kappa_2 -\gamma_2)}{\beta^{1/2}\left({x-a}\right)^{2} } \ L_2^{(-5/2)}\left(\frac{x-a}{\beta}\right) \right. \nonumber \\[2mm]
&-&  \left. \frac{(\kappa_3 -\gamma_3)}{\beta^{1/2}\left({x-a}\right)^{3} } \ L_3^{(-7/2)}\left(\frac{x-a}{\beta}\right) \ + \ \cdots\right].
\label{14aa}
\end{eqnarray}
%%%
If needed, one can use a relationship between the moments and the cumulants (Fa\`{a} di Bruno's formula~\cite{Lukacz})
to recast the expansion (\ref{14aa}) into more familiar language. For the Gram--Charlier A expansion various
formal convergence criteria exist (see, e.g.,~\cite{Wallaca:58}). In particular, the expansion for
nearly Gaussian equimeasurable PDFs ${\mathcal{F}(y)}$ converges quite rapidly and the series can be
truncated fairly quickly. Since in this case one needs fewer $\kappa_k$'s in order to determine the
information PDF $g(x)$, only EPs in the small neighborhood of the index $1$ will be needed. On the other hand,
the further the $\mathcal{F}({y})$ is from  Gaussian (e.g., heavy-tailed PDFs) the higher the orders of $\kappa_k$
are required to determine $g(x)$, and hence a wider neighborhood of the index $1$ will be needed for EPs.

%%%%%%%%%%%%%%%%%%%%%%%%%%%%%%%%%%%%%%%%%%%%%%%%%%%%%%%%%%%%%%%%%%%%%%%%%%%%%%%%%%%%%%%
\section{Example --- reconstruction theorem and (un)balanced cat state \label{sec6}}
%%%%%%%%%%%%%%%%%%%%%%%%%%%%%%%%%%%%%%%%%%%%%%%%%%%%%%%%%%%%%%%%%%%%%%%%%%%%%%%%%%%%%%%%
%*** discuss unbalanced cat states and related numerical analysis***

We now demonstrate an example of the reconstruction in the context of a quantum system. Specifically, we consider cat states that are
often considered in the foundations of quantum physics as well as in various applications, including
solid state physics~\cite{Cundiff2011} and quantum metrology~\cite{Knott2016}. The form of the state we
consider is $\ket{\psi}=\mathcal{N}(\ket{0} + \nu\ket{\alpha/\nu})$ where, $\mathcal{N}=[1+2\nu\exp(-\alpha^{2}/2\nu^{2}) +\nu^{2}]^{-1/2}$ is
the normalization factor, $\ket{0}$ is the vacuum state, $\nu\in\mathbb{R}$ a weighting factor and $\ket{\alpha}$ is the coherent state given by
\begin{align}
\ket{\alpha} \ = \ e^{-\alpha^{2}/2}\sum_{n=0}^{\infty}\frac{\alpha^{n}}{\sqrt{n!}} \ \! \ket{n}\, ,
\end{align}
(taking $\alpha\in\mathbb{R}$). For $\nu=1$ we refer to the state as a \textit{balanced cat state} (BCS) and for $\nu\neq 1$, as an \textit{unbalanced cat state} (UCS). Changing the basis of $\ket{\psi}$ to the eigenstates of the general quadrature operator
\begin{align}
\hat{Y}_{\theta}\ = \ \frac{1}{\sqrt{2}}\left(\hat{a}e^{-i\theta} + \hat{a}^{\dagger}e^{i\theta}\right)\, ,
\end{align}
where, $\hat{a}$ and $\hat{a}^{\dagger}$ are the creation and annihilation operators of the electromagnetic field, we find the PDF for the general quadrature variable $y_{\theta}$ to be
\begin{eqnarray}
\mathcal{F}(y_{\theta}) \ = \ \mathcal{N}^{2}\pi^{-\frac{1}{2}}e^{-y^{2}_{\theta}}\ \! \left|1+\nu\exp\left[ -\frac{\alpha^{2}}{\nu^{2}2}\left( 1 + e^{2i\theta} - 2\sqrt{2}e^{i\theta}\frac{\nu}{\alpha}y_{\theta} \right) \right]\right|^{2}\, ,
\label{eq1}
\end{eqnarray}
where $\mathcal{N}$ is the normalization constant. Setting $\theta=0$ and $\nu=1$ returns the PDF of the BCS for the position-like variable $y_{0}$. With this, the R\'{e}nyi EPs $N_{1-p}(\chi)$ are calculated and found to be constant across varying $p$. This is because $\mathcal{F}(y_{0})$ for the BCS is in fact a piecewise rearrangement of a Gaussian PDF (yet has an overall non-Gaussian structure) as depicted in Fig.~\ref{BCS}, thus $N_{1-p}(\chi)=\sigma^{2}$ for all $p$, where $\sigma^2$ is the variance of the `would be Gaussian'. Taking the reference PDF to be $G(x)=\mathcal{G}(x|a,\alpha,\beta)$, with $a=\log_{2}(2\pi\sigma^{2})/2$, $\alpha=1/2$ and $\beta=\log_{2}(e)$, it is evident that $(\kappa_{k}-\gamma_{k})=0$ for all $k\geq1$, and from the Gram--Charlier A series (\ref{12abc}), a perfect matching in the reconstruction is achieved. Furthermore, it can be shown that the variance of (\ref{eq1}) increases with $\alpha$, i.e. the variance increases as the peaks of the PDF diverge, which is in stark contrast to the R\'{e}nyi EPs which remain constant for increasing $\alpha$. This reveals the shortcomings of variance as a measure of uncertainty for non-Gaussian PDFs.

The peaks, located at $\mathcal{F}(y_{\theta})=2^{-a^{+}_{j}}$, where $j$ is an index labelling the distinct peaks, gives rise to sharp singularities in the target $g(x)$. With regard to the BCS position PDF, distributions of the conjugate parameter $\mathcal{F}(y_{\pi/2})$ distinguish
$\mathcal{F}(y_{0})$ from its equimeasurable Gaussian PDF and hence the R\'{e}nyi EPs also distinguish the different cases.
The number of available cumulants $k$ is computationally limited but as this grows, information about the singularities will
be recovered in the reconstruction. In the following, we show how the tail convergence and location of a singularity for
$g(x)$ can be reconstructed using $k=5$.
\begin{center}
	\includegraphics[scale=0.55]{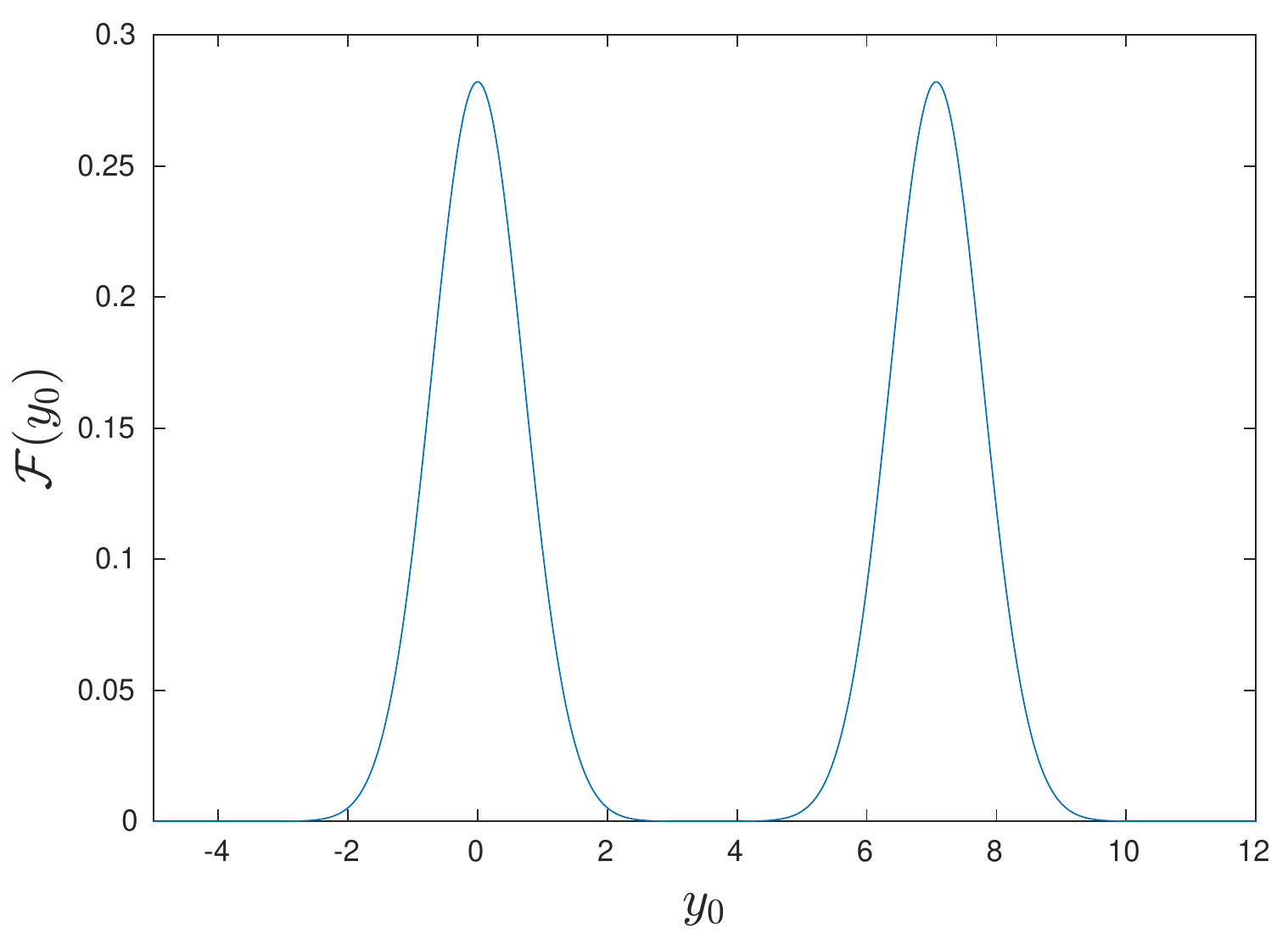}
	\captionof{figure}{Probability distribution function of a balanced cat state (BCS) for the quantum mechanical state's position-like quadrature variable with $\alpha=5$. This clearly displays an overall non-Gaussian structure, however, as this is a piecewise rearrangement of a Gaussian PDF for all $\alpha$ we have that  $N_{1-p}=\sigma^{2}$ for all $p$ and $\alpha$.  }
	\label{BCS}
\end{center}
\vspace{3mm}
%For (b) the first 4 cumulants have been used in the reconstruction of $g(x)$ and the exponential approximated to 2nd order. The value $a_{j}^{+}$ corresponds to the value of $x$ at the point of intersection with the $j$th distinct peak ($j=1$ not shown) of the $\mathcal{F}(y_{\pi/2})$
%
We consider the case of a UCS with $\nu=0.97$, $\alpha=10$ and we take $\theta=0$ in equation (\ref{eq1}) to find the PDF in the $y_{0}$ quadrature which is non-Gaussian for all piecewise rearrangements. As such, all REPs $N_{1-p}$ vary with $p$ and consequently all cumulants $\kappa_{k}$ carry information on $g(x)$. Here we choose to reconstruct the UCS information
distribution by means of the Edgeworth series~\cite{Wallaca:58} so that
\begin{equation}
\mbox{\hspace{-0.5mm}}g(x)=\exp\left[ n\sum_{j=2}^{\infty}(\kappa_{j} - \gamma_{j})\frac{(-1)^{j}}{j!}\frac{d^{j}}{dx^{j}}n^{-j/2} \right]G(x),
\end{equation}
where the reference PDF $G(x)$ is again the shifted gamma distribution. Using the Edgeworth series, the information PDF is approximated by expanding
in orders of $n$, which has the advantage over the Gram--Charlier~A expansion discussed above of bounding the errors of the approximation.
For the particular UCS of interest, expanding to order $n^{-3/2}$ reveals convergence toward the analytic form of the information
PDF shown as the target line in Fig.~\ref{UCS}. This shows that, for a given characteristic resolution, control over the first five
R\'{e}nyi EPs can be enough for a useful information scan of a quantum state with an underlying non-Gaussian PDF. In the example
shown in Fig.~\ref{UCS} we see that the information scan accurately predicts the tail behavior as well as the location of
the singularity, which corresponds to the second (lower) peak of $\mathcal{F}(y_{0})$.
\begin{center}
	\centering
	\includegraphics[scale=0.57]{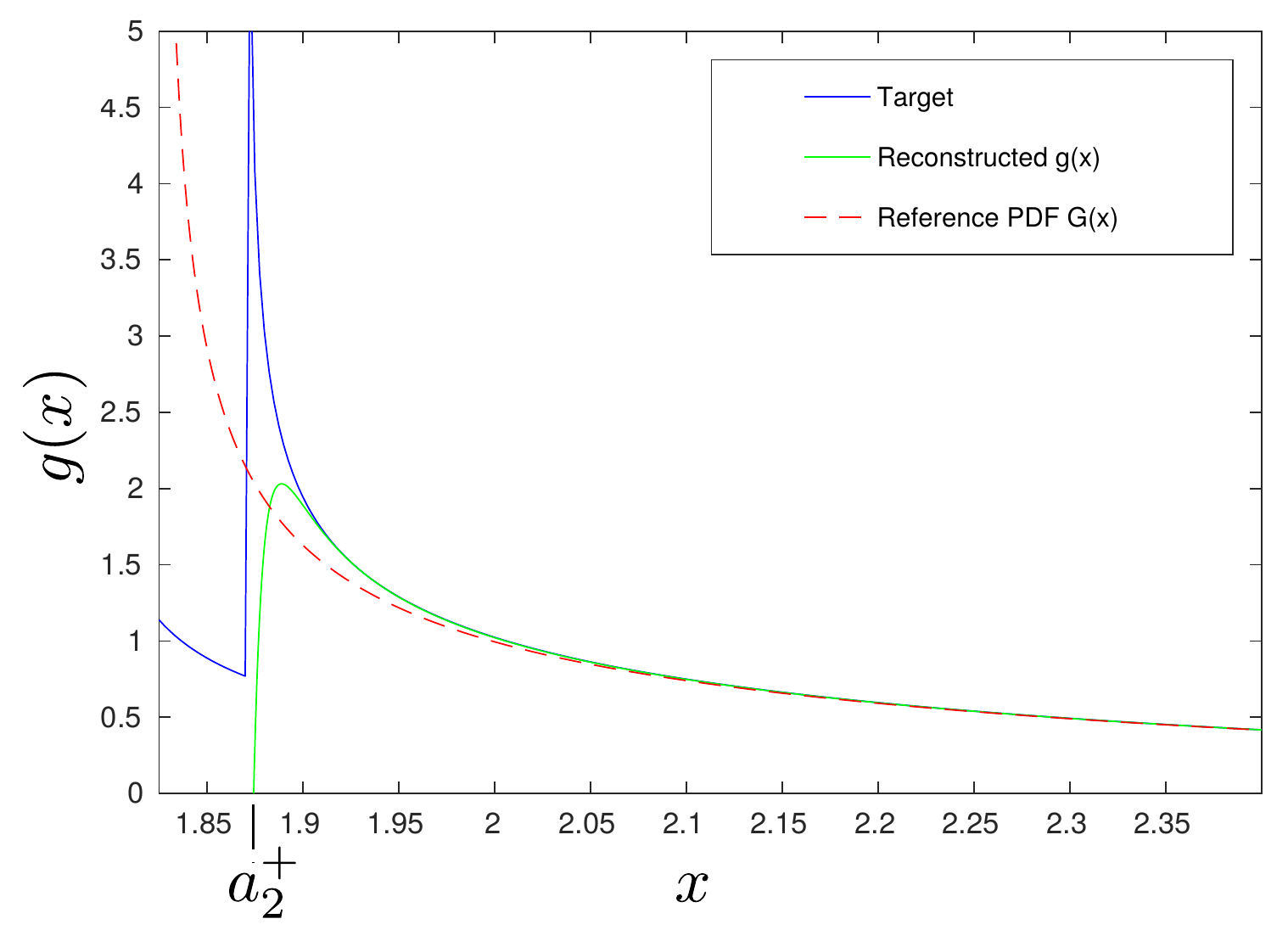}
	\captionof{figure}{Reconstructed information distribution of an unbalanced cat state with $\nu=0.97$ and $\alpha=10$. The Edgeworth expansion has been used here to order $n^{-3/2}$ requiring control of the first five REPs. Good convergence of the tail behaviour is evident as well as the location of the singularity corresponding to the second peak; $a_{2}^{+}$ corresponds to the value of $x$ at the point of intersection with the second (lower) peak of $\mathcal{F}(y_{0})$.}
	\label{UCS}
\end{center}
\vspace{3mm}
%

%%%%%%%%%%%%%%%%%%%%%%%%%%%%%%%%%%%%%%%%%%%%%%%%%%%%%%%%%%%%%%%%%%%%%%
\section{Entropy powers based on Tsallis entropy \label{sec6cd}}
%%%%%%%%%%%%%%%%%%%%%%%%%%%%%%%%%%%%%%%%%%%%%%%%%%%%%%%%%%%%%%%%%%%%%%
%
Let us now briefly comment on the entropy powers associated to yet another important
differential entropy, namely {\em Tsallis differential entropy}, which is defined as~\cite{tsallis2}
\begin{eqnarray}
{\mathcal{S}}_q(\mathcal{F}) \ = \ \frac{1}{(1-q)}\left[  \int_{\mathbb{R}^D}\left(\mathcal{F}^{q}({\boldsymbol{x}}) -  \mathcal{F}({\boldsymbol{x}})\right)d{\boldsymbol{x}} \right]\, ,
\end{eqnarray}
where, as before the PDF $\mathcal{F}({\boldsymbol{x}})$ is associated with a random vector $\{\mathcal{X}_i \}$ in $\mathbb{R}^D$.

Similarly to the RE case, the Tsallis entropy power $N_p^T(\mathcal{X})$ is defined as the
%~\cite{JDJ,Gardner02}:
solution of the equation
\begin{eqnarray}
{\mathcal{S}}_q\left( {\mathcal{X}} \right)
\ = \ {\mathcal{S}}_q^T\left(\sqrt{N_q^T(\mathcal{X})}\cdot
{\mathcal{Z}}^{_G}\right)\,
.\label{50.ccd}
\end{eqnarray}
The ensuing entropy power has not been studied in the literature yet but it can be easily  derived by observing that
the following scaling property for differential Tsallis entropy holds, namely
\begin{eqnarray}
{\mathcal{S}}_q(a \mathcal{X}) \ = \   {\mathcal{S}}_q(\mathcal{X}) \ \oplus_q \ \ln_q |a|^D\, ,
\label{51.bbc}
\end{eqnarray}
where $a \in \mathbb{R}$ and the $q$-deformed sum and logarithm are defined as~\cite{Tsallis-book}: $x \oplus_q  y  =  x  +  y  +  (1-q) x y$ and
$\ln_q x  =  (x^{1-q }- 1)/(1-q)$, respectively.
Relation (\ref{51.bbc}) results from the following chain of identities
\begin{eqnarray}
{\mathcal{S}}_q(a \mathcal{X}) \ &=& \  \frac{1}{1-q} \left[\int_{\mathbb{R}^D} d^D {\bf y}\left(\int_{\mathbb{R}^D} d^D {\bf x} \ \! \delta({\bf y} - a {\bf x}) {\mathcal{F}}({\boldsymbol{x}}) \right)^{\!q} -1 \right] \nonumber \\[2mm]
&=& \frac{1}{1-q} \left[|a|^{D(1-q)}\ \! \int_{\mathbb{R}^D} d^D {\bf y} \ \! {\mathcal{F}}^q({\boldsymbol{y}}) -1  \right]\nonumber \\[2mm]
&=& \ |a|^{D(1-q)}\ \! \left({\mathcal{S}}_q(\mathcal{X}) \ + \ \frac{1}{1-q} \right) \ - \ \frac{1}{1-q} \ = \ |a|^{D(1-q)}\ \!{\mathcal{S}}_q(\mathcal{X}) \ + \ \ln_q |a|^D
\nonumber \\[2mm]
&=& \ \left[(1-q) \ln_q |a|^D \ + \  1\right]{\mathcal{S}}_q(\mathcal{X}) \ + \ \ln_q |a|^D
\ = \ {\mathcal{S}}_q(\mathcal{X}) \ \oplus_q \ \ln_q |a|^D\, .
\label{scaling}
\end{eqnarray}
We can further use the simple fact that
\begin{eqnarray}
{\mathcal{S}}_q({\mathcal{Z}}_{G}) \ = \  \ln_q(2\pi q^{q'/q})^{D/2}\, .
\label{15.bc}
\end{eqnarray}
Here $q$ and $q'$ satisfy  $1/q + 1/q' =1$ with $q \in \mathbb{R}^+$. By combining (\ref{50.ccd}), (\ref{51.bbc}) and (\ref{15.bc}) we get
\begin{eqnarray}
{\mathcal{S}}_q(\mathcal{X}) \ = \ \ln_q(2\pi q^{q'/q})^{D/2} \ \oplus_q \ \ln_q (N_q^T)^D/2 \ = \ \ln_q(2\pi q^{q'/q}N_q^T)^{D/2}\, ,
\label{15.bcd}
\end{eqnarray}
where we have used the sum rule from the $q$-deformed calculus: $\ln_q x \oplus_q  \ln_q y = \ln_q xy$. Equation (\ref{15.bcd}) can be resolved for $N_p^T$ by employing the $q$-exponential, i.e., $e_q^x = [1 + (1-q)x]^{1/(1-q)}$, which (among others) satisfies the relation $e_q^{\ln_q x} = \ln_q (e_q^x) = x$. With this we have
\begin{eqnarray}
N_q^T(\mathcal{X}) \ = \ \frac{1}{2\pi} q^{-q'/q} \left[\exp_q \left({\mathcal{S}}_q(\mathcal{X})\right)\right]^{2/D} \ = \ \frac{1}{2\pi} q^{-q'/q} \exp_{1 - (1-q)D/2} \left(\frac{2}{D}\ \!{\mathcal{S}}_q(\mathcal{X})\right)\, .
\end{eqnarray}
In addition, when $q \rightarrow 1_+$ one has
\begin{eqnarray}
\lim_{q\rightarrow 1 }N_q^T(\mathcal{X}) \ = \ \frac{1}{2\pi e} \exp\left( \frac{2}{D} \mathcal{H}(\mathcal{X})\right) \ = \ N(\mathcal{X})\, ,
\end{eqnarray}
where $N(\mathcal{X})$ is the conventional Shannon entropy power and $\mathcal{H}(\mathcal{X})$ is the Shannon entropy~\cite{Shannon48}.

In connection with Tsallis EP we might notice one interesting fact, namely by starting from R\'{e}nyi's EP (considering RE in nats) we can write
\begin{eqnarray}
N_q (\mathcal{X})
\ &=& \ \frac{1}{2\pi} q^{-q'/q}
\exp\left(\frac{2}{D} \ \!{\mathcal{I}}_q({\mathcal{X}})\right) \ = \ \frac{1}{2\pi} q^{-q'/q} \left(\int d^D{\bf x} \ \! {\mathcal{F}}^q({\boldsymbol{x}})  \right)^{2/(D(1-q))} \nonumber \\[2mm]
&=& \ \frac{1}{2\pi} q^{-q'/q} \left[e_q^{{\mathcal{S}}_q^T(\mathcal{X})}\right]^{2/D} \ = \ N_q^T(\mathcal{X})\, .
\label{24aa}
\end{eqnarray}
Here we have used a simple identity
\begin{eqnarray}
\left(\int d^D{\bf x} \ \! {\mathcal{F}}^q({\boldsymbol{x}})  \right)^{{1}/{(1-q)}} \ = \ \left[(1-q){\mathcal{S}}_q^T({\mathcal{X}}) \ + \ 1   \right]^{1/(1-q)} \ = \ e_q^{{\mathcal{S}}_q^T(\mathcal{X})}\, .
\end{eqnarray}
So, we have obtained that R\'{e}nyi and Tsallis EPs coincide with each other. In particular, R\'{e}nyi's EPI (\ref{REPURaa}) can be equivalently written in the form
\begin{eqnarray}
%\mbox{\hspace{-3mm}}N_{1+t}({\mathcal{F}}^{(2)})N_{1+r}({\mathcal{F}}^{(1)})  \equiv
N_{p/2}^T({\mathcal{X}})N_{q/2}^T({\mathcal{Y}}) \ \geq  \  \frac{1}{16\pi^2}\, .
\label{TEPURaa}
\end{eqnarray}
Similarly, we could rephrase the generalized Stam inequality (\ref{stam.ineq.a}) and generalized isoperimetric inequality (\ref{2.12ac}) in terms of  Tsallis EPs. Though such inequalities are quite interesting from a mathematical point of view, it is not yet clear how they could be practically utilized in the estimation theory as there is no obvious operational meaning associated with Tsallis entropy (e.g., there is no coding theorem for Tsallis entropy). On the other hand, Tsallis entropy is important concept in the description of  entanglement~\cite{Wei:19}. For instance, Tsallis entropy of order $2$ (also known as linear entropy) directly quantifies state purity~\cite{Bengtsson}.

%
%%%%%%%%%%%%%%%%%%%%%%%%%%%%%%%%%%%%%%%%%%%%%%%%%%%%%%%%
\section{Conclusions\label{sec7}}
%%%%%%%%%%%%%%%%%%%%%%%%%%%%%%%%%%%%%%%%%%%%%%%%%%%%%%%%

%The advantage of these REPURs  over the conventional variance-based
%Robertson--Schr\"{o}dinger relation is that the right-hand side of the inequality is always
%independent of the considered quantum state and, in fact, it is a universal number
%for all EP orders.

In the first part of this paper we have introduce the notion of R\'{e}nyi's EP.
With quantum metrology applications in mind, we carried out our discussion in the framework of estimation theory.
In doing so, we have generalized the notion of Fisher information (FI) by using a R\'{e}nyi entropy
version of De~Bruijn's identity. The key role of the escort distribution in this context was highlighted.
With R\'{e}nyi's EP at hand, we proved the RE-based isoperimetric and Stam inequalities.
We have further clarified the role of R\'{e}nyi's EP by deriving (through the generalized Stam inequality)
a one-parameter family of R\'{e}nyi EP-based  quantum mechanical uncertainty
relations. Conventional variance-based URs of Robertson-Schr\"{o}dinger and Shannon differential entropy-based URs of
Hirschman or Bia{\l}ynicki-Birula  naturally appear as special cases in this hierarchy of URs.
Interestingly, we found that the Tsallis entropy-based EP coincided with R\'{e}nyi's EP provided that the order is the same.
This might open quite a new, hitherto unknown role for Tsallis entropy in estimation theory.

The second part of the paper was devoted to developing the application of R\'{e}nyi's EP for extracting quantum states
from incomplete data. This is of particular interest in various quantum metrology protocols. To
that end, we introduced the concepts of information distribution and showed how
cumulants of the information distribution can be obtained from knowledge of EPs of various orders.
With cumulants thus obtained, one can reconstruct the underlying information distribution in a process which we call
an information scan. A numerical implementation of this reconstruction procedure was technically realized via
Gram--Charlier A and Edgeworth expansion. For an explicit illustration of the information scan we used the
non-Gaussian quantum states --- (un)balanced cat states. In this case it was found that
control of the first five significant R\'{e}nyi EPs  gave enough information for a meaningful reconstruction of the
information PDF and brought about non-trivial information on the original balanced cat state PDF, such as asymptotic tail behavior
or the heights of the peaks.

Finally let us stress one more point. R\'{e}nyi EP-based  quantum mechanical uncertainty
relations (\ref{26bc}) basically represent a one-parameter class of inequalities that constrain higher-order cumulants
of state distributions for conjugate observables~\cite{JD:16}. In connection with this the following two questions are in order.
Assuming one is able to control R\'{e}nyi EPs of various orders:
i) how do such R\'{e}nyi EPs constrain the underlying state distribution and
ii) how do the ensuing REPURs restrict the state distributions of conjugate observables?
The first question was tackled in this paper in terms of the information distribution and
reconstruction theorem. The second question is more intriguing and has not yet been properly addressed.
Work along these lines is presently under investigation.

\vspace{6pt}

%%%%%%%%%%%%%%%%%%%%%%%%%%%%%%%%%%%%%%%%%%
%% optional
%\supplementary{The following are available online at \linksupplementary{s1}, Figure S1: title, Table S1: title, Video S1: title.}

% Only for the journal Methods and Protocols:
% If you wish to submit a video article, please do so with any other supplementary material.
% \supplementary{The following are available at \linksupplementary{s1}, Figure S1: title, Table S1: title, Video S1: title. A supporting video article is available at doi: link.}

%%%%%%%%%%%%%%%%%%%%%%%%%%%%%%%%%%%%%%%%%%
\authorcontributions{All authors jointly discussed, conceived and wrote the manuscript.}

%%%%%%%%%%%%%%%%%%%%%%%%%%%%%%%%%%%%%%%%%%

%%%%%%%%%%%%%%%%%%%%%%%%%%%%%%%%%%%%%%%%%%
\funding{P.J. and M.P. were  supported  by the Czech  Science  Foundation Grant No. 19-16066S. J.D. acknowledges support from DSTL and the UK EPSRC through the NQIT Quantum Technology Hub (EP/M013243/1) }
%%%%%%%%%%%%%%%%%%%%%%%%%%%%%%%%%%%%%%%%%%%%%%%%%%%%%%%%%%%%%%%%%%%%%%%%%%%%%%%%%%%%%%%%%%%%%%%%%%%%

%%%%%%%%%%%%%%%%%%%%%%%%%%%%%%%%%%%%%%%%%%
%%%%%%%%%%%%%%%%%%%%%%%%%%%%%%%%%%%%%%%%%%

%%%%%%%%%%%%%%%%%%%%%%%%%%%%%%%%%%%%%%%%%%
\conflictsofinterest{The authors declare no conflict of interest.}
%%%%%%%%%%%%%%%%%%%%%%%%%%%%%%%%%%%%%%%%%%
%% optional
\abbreviations{The following abbreviations are used in this manuscript:\\

\noindent
\begin{tabular}{@{}ll}
ITE & Information-theoretic entropy \\
UR & Uncertainty relation \\
RE & R\'{e}nyi entropy \\
TE & Tsallis entropy \\
REPUR & R\'{e}nyi entropy-power-based quantum uncertainty relation \\
QM & Quantum mechanics\\
EP & Entropy power\\
FI & Fisher information \\
PDF & Probability density function\\
EPI & Entropy power inequality \\
REP & R\'{e}nyi entropy power\\
BCS & Balanced cat state \\
UCS & Unbalanced cat state
\end{tabular}}

%%%%%%%%%%%%%%%%%%%%%%%%%%%%%%%%%%%%%%%%%%
%% optional
\appendixtitles{zes} % Leave argument "no" if all appendix headings stay EMPTY (then no dot is printed after "Appendix A"). If the appendix sections contain a heading then change the argument to "yes".
\appendix
%%%%%%%%%%%%%%%%%%%%%%%%%%%%%%%%%%%%%%%%%%%%%%%%%%%%%%%%%%%%%%%%%%%
\section*{Appendix A~\label{ap1}}
%%%%%%%%%%%%%%%%%%%%%%%%%%%%%%%%%%%%%%%%%%%%%%%%%%%%%%%%%%%%%%%%%%%

Here we provide an intuitive proof of the generalized  De~Bruijn identity.\\[-1mm]

\noindent {\em{Generalized De~Bruijn identity~I}:~} By denoting the PDF associated with a random vector $\{\mathcal{X}_i\}$ as $\mathcal{F}({\boldsymbol{x}})$ and the noise PDF as  $\mathcal{G}({\boldsymbol{z}})$, we might write the LHS of (\ref{DeB.2a}) as
\begin{eqnarray}
&&\mbox{\hspace{-20mm}}\frac{d}{d\epsilon} {\mathcal{I}}_q(\mathcal{X} + \sqrt{\epsilon}\ \! \mathcal{Z})|_{\epsilon = 0}\nonumber \\[2mm]
&=&  \left. \frac{1}{1-q} \frac{d}{d\epsilon} \log\left[\int_{\mathbb{R}^{D}} d{\boldsymbol{y}}  \left( \int_{\mathbb{R}^{D}} d{\boldsymbol{x}} \int_{\mathbb{R}^{D}} d{\boldsymbol{z}} \ \! \delta^{(D)}\left({\boldsymbol{y}}  - ({\boldsymbol{x}} + \sqrt{\epsilon}{\boldsymbol{z}})\right)   {\mathcal{F}}({\boldsymbol{x}})\mathcal{G}({\boldsymbol{z}}) \right)^{\!\!q} \ \!\right]\right|_{\epsilon = 0} \nonumber \\[2mm]
&=&  \left.\frac{1}{1-q} \frac{d}{d\epsilon} \log\left[\int_{\mathbb{R}^{D}} d{\boldsymbol{y}}  \left(\int_{\mathbb{R}^{D}} d{\boldsymbol{z}}  \ \! {\mathcal{F}}({\boldsymbol{y}} - \sqrt{\epsilon}{\boldsymbol{z}})  \mathcal{G}({\boldsymbol{z}}) \right)^{\!\!q}  \ \! \right]\right|_{\epsilon = 0} \nonumber \\[2mm]
&=& \left.\frac{1}{1-q} \frac{d}{d\epsilon} \log\left\{\int_{\mathbb{R}^{D}} d{\boldsymbol{y}}  \left[\int_{\mathbb{R}^{D}} d{\boldsymbol{z}}  \ \! \left({\mathcal{F}}({\boldsymbol{y}}) - \sqrt{\epsilon}{\boldsymbol{z}}_i \nabla_i{\mathcal{F}}({\boldsymbol{y}})\right. \right. \right. \right. \nonumber \\[2mm]
&+& \left.\left.\left. \left.   \frac{1}{2 }\epsilon {\boldsymbol{z}}_i{\boldsymbol{z}}_j \nabla_i\nabla_j{\mathcal{F}}({\boldsymbol{y}}) \ + \ {\mathcal{O}}({\epsilon}^{3/2})\right) \mathcal{G}({\boldsymbol{z}}) \right]^{\!q} \ \! \right\}\right|_{\epsilon = 0}\nonumber \\[2mm]
&=& \frac{q}{1-q} \left[\int_{\mathbb{R}^{D}} d{\boldsymbol{y}} \ \! \rho_q({\boldsymbol{y}}) {\boldsymbol{\Sigma}}_{ij} \frac{\nabla_i \nabla_j
{\mathcal{F}}({\boldsymbol{y}}) }{2{\mathcal{F}}({\boldsymbol{y}}) }\right] \ = \
 \frac{q}{2} \left[\int_{\mathbb{R}^{D}} d{\boldsymbol{y}} \ \! \rho_q({\boldsymbol{y}}) {\boldsymbol{\Sigma}}_{ij}  V_i({\boldsymbol{y}})V_j({\boldsymbol{y}})\right]\nonumber \\[2mm]
&=&  \frac{q}{2}\ \! \mbox{Tr}[\mbox{cov}_q({\boldsymbol{V}}) {\boldsymbol{\Sigma}}] \ = \ \frac{1}{2q}\ \! \mbox{Tr}[\mbox{cov}_q({\boldsymbol{V}}_q) {\boldsymbol{\Sigma}}] \ = \
\frac{1}{2q}\ \! \mbox{Tr}( {\mathbb{J}}_q{\boldsymbol{\Sigma}})\, .
\label{A.1.a}
\end{eqnarray}
The right-hand-side of (\ref{A.1.a}) can also be equivalently written as
\begin{eqnarray}
&&\mbox{\hspace{-12mm}}\frac{1}{2q}\mathbb{E}_q\left\{[(V_q)_i - \mathbb{E}_q((V_q)_i)]\Sigma_{ij}[(V_q)_j - \mathbb{E}_q((V_q)_j))]  \right\}\! ,\nonumber \\
&&\mbox{\hspace{10mm}}= \ \frac{1}{2q}\mathbb{E}\left\{[(\mathcal{Z}_i - \mathbb{E}(\mathcal{Z}_i)]({\mathbb{J}}_q)_{ij}(\mathcal{X})[\mathcal{Z}_j - \mathbb{E}(\mathcal{Z}_j)]  \right\}\! ,
\end{eqnarray}
where the mean $\mathbb{E}_q\{\ldots\}$ is performed with respect to the escort distribution $\rho_q$, while
$\mathbb{E}$ with respect to $\mathcal{G}$ distribution.

We note in passing that the conventional De~Bruijn's identity (\ref{DBI.5a}) emerges as a special case when $q \rightarrow 1$.  For the Gaussian noise
vector we can generalize the previous derivation in the following way:\\[0mm]

\noindent {\em{Generalized De~Bruijn's identity~II}:~} Let $\{\mathcal{X}_i\}$ be a random vector in ${\mathbb{R}}^D$ with the PDF $\mathcal{F}({\boldsymbol{x}})$
and let $\{\mathcal{Z}_i\}$ be an independent Gaussian noise vector with the zero mean and covariance matrix ${\boldsymbol{\Sigma}} = \mbox{cov}(\mathcal{Z}^{_G})$, then
%
%\begin{widetext}
\begin{eqnarray}
\frac{d}{d{{\Sigma}}_{ij}} {\mathcal{I}}_q(\mathcal{X} +  \mathcal{Z}^{_G})|_{_{{\boldsymbol{\Sigma}}= 0}}
&=& \frac{q}{1-q} \left[\int_{\mathbb{R}^{D}} \!\! d{\boldsymbol{y}}
\ \! \rho_q({\boldsymbol{y}})  \frac{\nabla_i \nabla_j
{\mathcal{F}}({\boldsymbol{y}}) }{2{\mathcal{F}}({\boldsymbol{y}}) }\right] \nonumber \\[2mm]
&=& \frac{1}{2q} \left[\int_{\mathbb{R}^{D}} \!\! d{\boldsymbol{y}}
\ \! \rho_q({\boldsymbol{y}})  ({{V}}_q)_i({{V}}_q)_j \right] \ = \ \frac{1}{2q}\ \! ({\mathbb{J}}_q)_{ij}\, .
\label{A.3.a}
\end{eqnarray}
%\end{widetext}
%
The right-hand-side is equivalent to
\begin{eqnarray}
\frac{1}{2q}\mathbb{E}_q\left\{[(V_q)_i - \mathbb{E}_q((V_q)_i)][(V_q)_j - \mathbb{E}_q((V_q)_j))]  \right\}\! .
\end{eqnarray}
To prove the identity (\ref{A.3.a}), we might follow the same line of reasonings as in  (\ref{A.1.a}). The only difference is that while in (\ref{A.1.a}) we had a small parameter $\epsilon$ in which one could expand to all orders of correlation functions and easily perform differentiation and limit $\epsilon \rightarrow 0$ for any noise distribution (with zero mean), the same procedure can not be done in the present context for a generic noise distribution. In fact, only the Gaussian distribution has the property that the higher-order correlation functions and their derivatives with respect to ${\boldsymbol{\Sigma}}_{ij}$ are small when ${\boldsymbol{\Sigma}}$ is small. The latter is a consequence of the Marcinkiewicz theorem~\cite{Marcinkiewicz:39}.

%%%%%%%%%%%%%%%%%%%%%%%%%%%%%%%%%%%%%%%%%%%%%%%%%%%%%%%%%%%%%%%%%%%
\section*{Appendix B~\label{ap2}}
%%%%%%%%%%%%%%%%%%%%%%%%%%%%%%%%%%%%%%%%%%%%%%%%%%%%%%%%%%%%%%%%%%%

Here we prove the {\em Generalized isoperimetric inequality} from Section~\ref{sec2}.
The starting point is the {\em entropy-power inequality} (EPI)~\cite{JDJ}:
Let $\mathcal{X}_{1}$ and $\mathcal{X}_{2}$ be two independent
continuous random vectors in $\mathbb{R}^{D}$ with probability
densities ${\mathcal{F}}^{(1)} \in \ell^{q}({\mathbb{R}^{D}})$ and
${\mathcal{F}}^{(2)} \in \ell^{p}({\mathbb{R}^{D}})$, respectively.
Suppose further that $\lambda \in (0,1)$ and $r>1$, and let
\begin{eqnarray}
q \ = \ \frac{r}{(1-\lambda) + \lambda r}\, , \;\;\;\; p =
\frac{r}{\lambda + (1-\lambda) r}\, ,
\label{B39aa}
\end{eqnarray}
then the following inequality holds:
\begin{eqnarray}
{N}_{r}(\mathcal{X}_{1}+\mathcal{X}_{2}) \ \geq \ \left(\frac{{N}_{q}
(\mathcal{X}_{1})}{1-\lambda} \right)^{1-\lambda} \left(\frac{{N}_{p}
(\mathcal{X}_{2})}{\lambda} \right)^{\lambda}. \label{3.1.0a}
\end{eqnarray}
Let us now consider a Gaussian noise vector $\mathcal{Z}^{_G}$ (independent of $\mathcal{X}$)
with zero mean and covariance matrix ${\boldsymbol{\Sigma}}$.
Within this setting we can write the following EPIs
\begin{eqnarray}
{N}_{r}(\mathcal{X} + {\mathcal{Z}}^{_G})  \ \geq \ \epsilon^{\lambda}\left(\frac{1}{1-\lambda}\right)^{1-\lambda} \left(\frac{1}{\lambda}  \right)^{\lambda} [N_q(\mathcal{X})]^{1-\lambda} \, ,   \label{B38a} \\[2mm]
{N}_{r}(\mathcal{X} + {\mathcal{Z}}^{_G})
\ \geq \ \epsilon^{1-\lambda}\left(\frac{1}{1-\lambda}\right)^{1-\lambda} \left(\frac{1}{\lambda}  \right)^{\lambda} [N_p(\mathcal{X})]^{\lambda} \, ,
\label{B39a}
\end{eqnarray}
with $\epsilon \equiv \det(\boldsymbol{\Sigma})^{1/D}$. Here we have used the simple fact that ${N}_{r}({\mathcal{Z}}^{_G}) = \det(\boldsymbol{\Sigma})^{1/D}$, irrespective of the value of $r$.

Let us now fix $r$  and maximize the RHS of inequality (\ref{B38a})  with respect to $\lambda$ and $q$ provided we keep the constraint condition (\ref{B39aa}).
This yields the condition extremum
\begin{eqnarray}
\lambda  \ = \  \frac{\epsilon}{N_q(\mathcal{X}) }\ \! \exp\left[q(1-q)\frac{d \log{N_q(\mathcal{X})}}{d q}\right] \ + \  \mathcal{O}(\epsilon^2)\, .
\end{eqnarray}
With this, $q$ turns out to be
\begin{eqnarray}
 q \ = \ r \ + \ \frac{\epsilon (1-r)r}{N_r(\mathcal{X}) }\ \! \exp\left[(1-r)r\frac{d \log{N_r(\mathcal{X})}}{d r}\right]  \ + \   \mathcal{O}(\epsilon^2)\, ,
\label{B43bc}
\end{eqnarray}
which in the limit  $\epsilon \rightarrow 0$ reduces to $q= r \geq 1$. The latter implies that $p = 1$.
The result (\ref{B43bc}) implies that the RHS of (\ref{B38a}) reads
\begin{eqnarray}
N_q(\mathcal{X}) \ + \ \epsilon \ \! \exp\left[(1-r)r\frac{d \log{N_r(\mathcal{X})}}{d r}\right] \left[1 - (1-r)r \frac{d \log{N_r(\mathcal{X})}}{d r}\right] \ + \ \mathcal{O}(\epsilon^2)\, .
\end{eqnarray}
Should we have started with the $p$ index, we would arrive at an analogous conclusion. %Namely, we would find that
%the maximizer is for $\lambda = N_q(\mathcal{X})/(N_q(\mathcal{X}) + \epsilon)$, the RHS of~(\ref{B39a}) turns out to be
%$N_p(\mathcal{X}) + \epsilon + \mathcal{O}(\epsilon^2)$ and
%%
%\begin{eqnarray}
%p \ = \ \frac{r(N_p(\mathcal{X}) + \epsilon)}{N_q(\mathcal{X}) + \epsilon r}\, .
%\end{eqnarray}
%
To proceed, we stick, without loss of generality, to the inequality (\ref{B38a}). This implies that
\begin{eqnarray}
{N}_{r}(\mathcal{X} +{\mathcal{Z}}^{_G}) & \geq & N_q(\mathcal{X}) \ + \ \epsilon \ \! \exp\left[(1-r)r\frac{d \log{N_r(\mathcal{X})}}{d r}\right] \left[1 - (1-r)r \frac{d \log{N_r(\mathcal{X})}}{d r}\right] \ + \ \mathcal{O}(\epsilon^2)\nonumber \\[2mm]
& = & N_r(\mathcal{X})  \ + \  [N_q(\mathcal{X}) - N_r(\mathcal{X})]  \nonumber \\[2mm]
&+& \epsilon \ \! \exp\left[(1-r)r\frac{d \log{N_r(\mathcal{X})}}{d r}\right] \left[1 - (1-r)r \frac{d \log{N_r(\mathcal{X})}}{d r}\right] \ + \ \mathcal{O}(\epsilon^2) \nonumber \\[2mm]
& \geq & N_r(\mathcal{X}) \ + \ \epsilon \ \! \exp\left[(1-r)r\frac{d \log{N_r(\mathcal{X})}}{d r}\right]  \ + \ \mathcal{O}(\epsilon^2)  \, .
\label{B47}
\end{eqnarray}
%
%where on the last line we have used the fact that [cf. Eq.~(\ref{B43bc})]
%%
%\begin{eqnarray}
%N_q(\mathcal{X}) - N_r(\mathcal{X}) & = &  \epsilon r(1-r) \frac{d \log N_{r}(\mathcal{X})}{d r}  +  \mathcal{O}(\epsilon^2) \nonumber \\
%& = &
%\end{eqnarray}
%
To proceed, we  employ the identity $\log{N_r(\mathcal{X})} = 2/D [{\mathcal{I}}_r(\mathcal{X}) - {\mathcal{I}}_r({\mathcal{Z}}^{_G}_{\ide})]$ with ${\mathcal{Z}}^{_G}_{\ide}$ representing a Gaussian random vector with zero mean
and {\em unit} covariance matrix, and the fact that ${\mathcal{I}}_r$ is monotonously decreasing function of $r$, i.e., $d {\mathcal{I}}_r/ dr \leq 0$ (see, e.g., Ref.~\cite{Renyi1970}).
With this we have
\begin{eqnarray}
\exp\left[(1-r)r\frac{d \log{N_r(\mathcal{X})}}{d r}\right] \ &\geq& \  \exp\left[\frac{2 (r-1)r }{D} \ \! \frac{d {\mathcal{I}}_r({\mathcal{Z}}^{_G}_{\ide})}{d r}  \right] \ = \ \exp\left[ (r-1)r  \frac{d}{d r} \!\left(\frac{1}{r-1}\log r \right) \right] \nonumber \\[2mm]
&=& \ e r^{r/(r-1)} \ \geq \ \frac{e^2}{r}\, .
\end{eqnarray}
Consequently, Eq.~(\ref{B47}) can be rewritten as
\begin{eqnarray}
&&\mbox{\hspace{-10mm}} \frac{{N}_{r}(\mathcal{X} +{\mathcal{Z}}^{_G}) \ - \ N_q(\mathcal{X})}{\Sigma_{ij}} \ \geq \  \frac{\epsilon}{\Sigma_{ij}}\frac{e^2}{r} \ + \
\mathcal{O}(\epsilon^2/\Sigma_{ij})\, .
\label{B48bc}
\end{eqnarray}
At this stage we are interested in the  ${{\Sigma}_{ij}}\rightarrow {{0}}$ limit. In order to find the ensuing leading order behavior of ${\epsilon}/{\Sigma_{ij}}$ we can use L'Hospital's rule, namely
\begin{eqnarray}
\frac{\epsilon}{\Sigma_{ij}} \ = \ \frac{d \epsilon }{d \Sigma_{ij}} \ = \ \frac{d}{d \Sigma_{ij}}\exp\left[\frac{1}{D}\mbox{Tr} (\log {\boldsymbol{\Sigma}})  \right]\ = \ \frac{\epsilon}{D} ({\boldsymbol{\Sigma}}^{-1})_{ij}\, .
\end{eqnarray}
Now we neglect the sub-leading term of order $\mathcal{O}(\epsilon^2/\Sigma_{ij})$ in (\ref{B48bc}) and take $\det(\ldots)^{1/D}$ on both sides.
This gives
\begin{eqnarray}
\left.\det\left(\frac{d {N}_{r}(\mathcal{X} +{\mathcal{Z}}^{_G})}{d\Sigma_{ij}}  \right)^{1/D}\right|_{{\boldsymbol{\Sigma}}= 0} \ = \ \frac{1}{rD} {N}_{r}(\mathcal{X}) [\det({\mathbb{J}}_r(\mathcal{X}))]^{1/D} \ \geq \ \frac{e^2}{rD} \ \geq \ \frac{1}{rD} \, ,
\end{eqnarray}
or equivalently
\begin{eqnarray}
{N}_{r}(\mathcal{X}) [\det({\mathbb{J}}_r(\mathcal{X}))]^{1/D} \ \geq \ 1\, .
\end{eqnarray}
At this stage we can use the inequality of arithmetic and geometric means to write
(note that ${\mathbb{J}}_r = \mbox{cov}_r({\boldsymbol{V}}_r)$ is a positive semi-definite matrix)
\begin{eqnarray}
\frac{1}{D}\mbox{Tr}({\mathbb{J}}_r(\mathcal{X})) \ \geq \ [\det({\mathbb{J}}_r(\mathcal{X}))]^{1/D} \, .
\label{app.B.58}
\end{eqnarray}
Consequently we have
\begin{eqnarray}
\frac{1}{D} {N}_{r}(\mathcal{X})\mbox{Tr}({\mathbb{J}}_r(\mathcal{X}))  \ = \  \frac{1}{D} {N}_{r}(\mathcal{X}){J}_r(\mathcal{X}) \ \geq \  {N}_{r}(\mathcal{X}) [\det({\mathbb{J}}_r(\mathcal{X}))]^{1/D} \ \geq \ 1\, ,
\end{eqnarray}
as stated in Eq.~(\ref{2.12ac}).

%%%%%%%%%%%%%%%%%%%%%%%%%%%%%%%%%%%%%%%%%%%%%%%%%%%%%%%%%%%%%%%%%%%
\section*{Appendix C~\label{ap3}}
%%%%%%%%%%%%%%%%%%%%%%%%%%%%%%%%%%%%%%%%%%%%%%%%%%%%%%%%%%%%%%%%%%%

In this appendix we prove the {\em Generalized Stam inequality} from Section~\ref{sec2}. We start with the defining relation (\ref{3.1.0e}), i.e.
\begin{eqnarray}
N_{q}(\mathcal{Y}) \ = \ \frac{1}{2\pi} q^{1/(1-q)} |\! | \mathcal{G} |\! |_q^{2q/[(1-q)D]}\, ,
\label{app.C.60}
\end{eqnarray}
and consider $q\in [1/2, 1]$ so that $q/(1-q) > 0$. For the $q$-norm  we can write
\begin{eqnarray}
|\! | \mathcal{G} |\! |_q \ = \ \left(\int_{\mathbb{R}^D} d{\boldsymbol{y}} \! \ |\psi_{_{\mathcal{G}}}({\boldsymbol{y}} )|^{2q}\right)^{1/q} \ = \ |\! | \psi_{_{\mathcal{G}}} |\! |_{2q}^2 \ \geq \ |\! | \hat{\psi}_{_{\mathcal{G}}} |\! |_{2r}^2 \ = \ |\! | {\psi}_{_{\mathcal{F}}} |\! |_{2r}^2 \ = \ |\! | \mathcal{F} |\! |_r\, .
\label{app.C.61}
\end{eqnarray}
Here $2r$ and $2q$ are H\"{o}lder conjugates so that $r \in [1, \infty]$. The inequality employed is due
to the Hausdorff--Young inequality (which in turn is a simple consequence of the H\"{o}lder inequality~\cite{JDJ}). We further have
\begin{eqnarray}
|\! | \mathcal{F} |\! |_r  &=&
 \left(\int_{\mathbb{R}^D} d{\boldsymbol{x}} \! \ |\psi_{_{\mathcal{F}}}({\boldsymbol{x}} )|^{2r}\right)^{1/r}
 \geq  \left| \int_{\mathbb{R}^D} d{\boldsymbol{x}} \! \ |\psi_{_{\mathcal{F}}}({\boldsymbol{x}} )|^{2r}\frac{\nabla_i \nabla_i
 \ \! e^{i \boldsymbol{a}\cdot \boldsymbol{x} }}{a_i^2} \right|^{1/r}
%=  \ \left|  2r \! \int_{\mathbb{R}^D} d{\bf x} \! \ |\psi_{_{\mathcal{F}}}({\bf x} )|^{2r-1}
%\nabla_i |\psi_{_{\mathcal{F}}}({\bf x} )| \frac{ e^{i \bf{a}\cdot\bf{x} }}{a_i} \right|^{1/r}
\nonumber \\[2mm]
%&=& \   \left| r \! \int_{\mathbb{R}^D} d{\boldsymbol{x}} \! \ \rho_r({\boldsymbol{x}} ) {{V}}_i({\boldsymbol{x}} ) \nabla_i  e^{i  \boldsymbol{a}\cdot \boldsymbol{x} } \right|^{1/r}\! \frac{\left(\int_{\mathbb{R}^D} d{\boldsymbol{x}} \! \
%|\psi_{_{\mathcal{F}}}({\boldsymbol{x}} )|^{2r}\right)^{1/r}}{|a_i|^{2/r}}\, ,
%\nonumber \\[2mm]
&=&  \left| r \int_{\mathbb{R}^D} d{\boldsymbol{x}} \! \ \left[ (r-1) \rho_r({\boldsymbol{x}} ) {{V}}_i({\boldsymbol{x}} ) {{V}}_i({\boldsymbol{x}} ) + \rho_r({\boldsymbol{x}} ) \frac{\nabla_i  \nabla_i \ \! \mathcal{F}({\boldsymbol{x}} ) }{\mathcal{F}({\boldsymbol{x}} )}  \right] e^{i  \boldsymbol{a}\cdot \boldsymbol{x} }   \right|^{1/r}\!  \frac{\left(\int_{\mathbb{R}^D} d{\boldsymbol{x}} \! \
|\psi_{_{\mathcal{F}}}({\boldsymbol{x}} )|^{2r}\right)^{1/r}}{a_i^{2/r}}
\nonumber \\[2mm]
& \geq &  \left| r \int_{\mathbb{R}^D} d{\boldsymbol{x}} \! \ \left[ (r-1) \rho_r({\boldsymbol{x}} ) {{V}}_i({\boldsymbol{x}} ) {{V}}_i({\boldsymbol{x}} ) + \rho_r({\boldsymbol{x}} ) \frac{\nabla_i  \nabla_i \ \! \mathcal{F}({\boldsymbol{x}} ) }{\mathcal{F}({\boldsymbol{x}} )}  \right] \cos(\boldsymbol{a}\cdot \boldsymbol{x} )  \right|^{1/r}\!  \frac{\left(\int_{\mathbb{R}^D} d{\boldsymbol{x}} \! \
|\psi_{_{\mathcal{F}}}({\boldsymbol{x}} )|^{2r}\right)^{1/r}}{a_i^{2/r}}
\nonumber \\[2mm]
& \geq &  \left| r \int_{V_D} d{\boldsymbol{x}} \! \   \rho_r({\boldsymbol{x}} ) \frac{\nabla_i  \nabla_i \ \! \mathcal{F}({\boldsymbol{x}} ) }{\mathcal{F}({\boldsymbol{x}} )}  \ \! \cos(\boldsymbol{a}\cdot \boldsymbol{x} )  \right|^{1/r}\!  \frac{\left(\int_{\mathbb{R}^D} d{\boldsymbol{x}} \! \
|\psi_{_{\mathcal{F}}}({\boldsymbol{x}} )|^{2r}\right)^{1/r}}{a_i^{2/r}}\, ,
%\geq  \left| r \! \int_{\mathbb{R}^D} d{\bf x} \! \ \rho_r({\bf x} ) {\boldsymbol{V}}_i({\bf x} ) e^{i \bf{a}\cdot\bf{x} }
% \right|^{D/r}\! \frac{\left(\int_{\mathbb{R}^D} d{\bf x} \! \ |\psi_{_{\mathcal{F}}}({\bf x} )|^{2r}\right)^{1/r}}{|a_i|^{D/r}} ,
\label{App.C.2}
\end{eqnarray}
where ${\boldsymbol{a}} \in \mathbb{R}^D$ is an arbitrary ${\boldsymbol{x}}$-independent vector, $\nabla_i \equiv \partial/\partial x_i$ and $V_D$ denotes a regularized volume of ${\mathbb{R}^D}$ --- $D$-dimensional ball of a very large (but finite) radius $R$.
In the first line of (\ref{App.C.2}) we have employed the triangle inequality $|{\mathbb{E}}_r\left( e^{i \boldsymbol{a}\cdot \boldsymbol{x}} \right) | \leq 1$ (with equality if and only if
$\boldsymbol{a} = \boldsymbol{0}$), namely
\begin{eqnarray}
\left| \int_{\mathbb{R}^D} d{\boldsymbol{x}} \! \ |\psi_{_{\mathcal{F}}}({\boldsymbol{x}} )|^{2r} e^{i \boldsymbol{a}\cdot \boldsymbol{x} } \right|\ = \ \left| \int_{\mathbb{R}^D} d{\boldsymbol{x}}
\! \ \rho_r({\boldsymbol{x}}) \ \! e^{i \boldsymbol{a}\cdot \boldsymbol{x} }   \right|  \int_{\mathbb{R}^D} d{\boldsymbol{x}} \! \ |\psi_{_{\mathcal{F}}}({\boldsymbol{x}} )|^{2r} \ \leq \
\int_{\mathbb{R}^D} d{\boldsymbol{x}} \! \ |\psi_{_{\mathcal{F}}}({\boldsymbol{x}} )|^{2r}.
\end{eqnarray}
The inequality in the last line  holds for $a_i = \pi/(2R)$ (for all  $i$)  since in this case  $\cos(\boldsymbol{a}\cdot \boldsymbol{x} ) \geq 0$  for all $\boldsymbol{x}$ from the $D$-dimensional ball.  In this case one may further estimate the integral from below by neglecting the positive integrand
$(r-1) \rho_r({\boldsymbol{x}} ) [{{V}}_i({\boldsymbol{x}} )]^2 $.

Note that (\ref{App.C.2}) implies
\begin{eqnarray}
\frac{ r \left|{\mathbb{E}}_r\left[{\mathcal{F}}^{-1} \nabla_i  \nabla_i \ \! \mathcal{F}  \cos({\boldsymbol{a}\cdot \boldsymbol{x} }) \right]\right|}{ a_i^2} \ \leq \ 1\, ,
\label{C.64.b}
\end{eqnarray}
with equality if and only if $\boldsymbol{a} \rightarrow \boldsymbol{0}$ (to see this one should apply L'Hospital's rule). Eq.~(\ref{C.64.b})
allows to write
\begin{eqnarray}
\mbox{\hspace{-4mm}}|\! | \mathcal{F} |\! |_r \ &\geq & \frac{ r^{\gamma} \left|{\mathbb{E}}_r\left[{\mathcal{F}}^{-1} \nabla_i  \nabla_i \ \! \mathcal{F}  \cos({\boldsymbol{a}\cdot \boldsymbol{x} }) \right]\right|^{\gamma}}{ a_i^{2\gamma}}\left(\int_{\mathbb{R}^D}
d{\boldsymbol{x}} \! \ |\psi_{_{\mathcal{F}}}({\boldsymbol{x}} )|^{2r}\right)^{1/r} \nonumber \\[2mm]
&\geq& \frac{ r^{\gamma} \left|{\mathbb{E}}_r\left[{\mathcal{F}}^{-1} \nabla_i  \nabla_i \ \! \mathcal{F}  \cos({\boldsymbol{a}\cdot \boldsymbol{x} }) \right]\right|^{\gamma}}{ a_i^{2\gamma}} \frac{1}{V_D^{1-1/r}}
\nonumber \\[2mm]
&=& \ \frac{ r^{\gamma} \left|{\mathbb{E}}_r\left[{\mathcal{F}}^{-1} \nabla_i  \nabla_i \ \! \mathcal{F}  \cos({\boldsymbol{a}\cdot \boldsymbol{x} }) \right]\right|^{\gamma}}{ a_i^{2\gamma}}\frac{1}{C_D^{1-1/r} R^{D-D/r}}
%\nonumber \\[2mm]
%&\geq& \ \frac{|{\mathbb{E}}_r[({V}_r)_i ({{V}}_r)_i e^{i  \boldsymbol{a}\cdot \boldsymbol{x} } ]|^{\gamma}}{|a_i|^{\gamma}|({{V}}_r)_i|_{\rm{max}}^{\gamma}}
%\frac{1}{C_D^{1-1/r} R^{D-D/r}}
\, ,
\label{C.65}
\end{eqnarray}
%In the second inequality we repeated the first inequality $D$ times.
%
where $\gamma > 0$ is some as yet unspecified constant and $C_D = \pi^{D/2}/\Gamma(D/2 + 1)$. In deriving (\ref{C.65}) we have
used the H\"{o}lder inequality
\begin{eqnarray}
1 \ &=& \ \left(\int_{\mathbb{R}^D} d{ \boldsymbol{x}} \! \ 1 \cdot |\psi_{_{\mathcal{F}}}({\boldsymbol{x}} )|^{2}\right)\ \leq  \ \left(\int_{\mathbb{R}^D} d{\boldsymbol{x}} \! \ 1^{r'}\right)^{1/r'}
 \left(\int_{\mathbb{R}^D} d{\boldsymbol{x}} \! \ |\psi_{_{\mathcal{F}}}({\boldsymbol{x}} )|^{2r}\right)^{1/r}\nonumber \\[2mm]
&=& \ V_D^{1-1/r}  \left(\int_{\mathbb{R}^D} d{\boldsymbol{x}} \! \ |\psi_{_{\mathcal{F}}}({\boldsymbol{x}} )|^{2r}\right)^{1/r},
\end{eqnarray}
Here [and also in (\ref{App.C.2}) and (\ref{C.65})] $V_D = C_D R^D$ denotes the regularizated volume of ${\mathbb{R}^D}$.

%We can further estimate (\ref{App.C.2}) from below by realizing that
%
%\begin{eqnarray}
%\left| \int_{\mathbb{R}^D} d{\bf x} \! \ \rho_r({\bf x} ) {\boldsymbol{V}}_i({\bf x} ) e^{i  \bf{a}\cdot\bf{x} } \right|^{1/r} \ &=& \ \frac{\left| \int_{\mathbb{R}^D} d{\bf x} \! \ \rho_r({\bf x} ) {\boldsymbol{V}}_i({\bf x} ) e^{i \bf{a}\cdot\bf{x} } \right|}{\left| \int_{\mathbb{R}^D} d{\bf x} \! \ \rho_r({\bf x} ) {\boldsymbol{V}}_i({\bf x} ) e^{i \bf{a}\cdot\bf{x} } \right|^{1-1/r}} \ \geq \ \frac{\left(\int_{\mathbb{R}^D} d{\bf x} \! \ \rho_r({\bf x} ) |{\boldsymbol{V}}_i({\bf x} )|^2\right)^{1/2}}{\left| \int_{\mathbb{R}^D} d{\bf x} \! \ \rho_r({\bf x} ) {\boldsymbol{V}}_i({\bf x} ) e^{i  \bf{a}\cdot\bf{x} } \right|^{1-1/r}}\nonumber \\[2mm]
%&\geq& \ \left(\int_{\mathbb{R}^D} d{\bf x} \! \ \rho_r({\bf x} ) |{\boldsymbol{V}}_i({\bf x} )|^2\right)^{1/2} \left(\frac{|a_i|}{r}\right)^{1/r-1} .
%\end{eqnarray}
%
%Here, the first inequality is due to Jensen's inequality for convex functions and the
%second comes from  Eq.~(\ref{App.C.2}).

As already mentioned, the best estimate of the inequality (\ref{C.65}) is obtained for $\boldsymbol{a} \rightarrow \boldsymbol{0}$.
As we have seen $a_i$ goes to zero  as $\pi/(2R)$ which allows to chose the  constant $\gamma$
so that the denominator in (\ref{C.65}) stays finite in the limit $R\rightarrow \infty$. This implies that $\gamma = D/2 - D/(2r)$.
Consequently,  (\ref{C.65}) acquires in the large $R$ limit the form
\begin{eqnarray}
|\! | \mathcal{F} |\! |_r \ \geq \ \frac{[4 (r-1)/r]^{D/2 -D/2r} \ \! [\Gamma(D/2 + 1)]^{1-1/r} }{\pi^{3D/2 -3D/2r}} \!\ [(\mathbb{J}_r)_{ii}(\mathcal{X})]^{D/2 -D/2r}\, ,
\end{eqnarray}
%
%This ensures that
%$e^{i  \boldsymbol{a}\cdot \boldsymbol{x} } \approx 1$ in ${\mathbb{E}}_r[\cdots]$ (in the regularized volume $V_D$) and we can write that
%${\mathbb{E}}_r[({V}_r)_i ({{V}}_r)_i e^{i  \boldsymbol{a}\cdot \boldsymbol{x} } ] \approx  (\mathbb{J}_r)_{ii}(\mathcal{X})$.
%
With this we can write [see Eqs.~(\ref{app.C.60})-(\ref{app.C.61})]
\begin{eqnarray}
N_{q}(\mathcal{Y}) \ \geq \
%\frac{1}{2\pi}\ \! r^{2r/(r-1)} q^{1/(1-q)} \left(\int_{\mathbb{R}^D} d{\bf x} \! \ \frac{\rho_r({\bf x} )}{|a_i|^{2}} |{\boldsymbol{V}}_i({\bf x} )|^2\right)^{q/(1-q)} \left(\int_{\mathbb{R}^D} d{\bf x} \! \ |\psi_{_{\mathcal{F}}}({\bf x} )|^{2r}\right)^{2/[(r-1)D]}\nonumber\\[2mm]
%&=& \
\frac{1}{(2\pi)^2}\ \!  q^{1/(1-q)}\ \! [(\mathbb{J}_r)_{ii}(\mathcal{X})]  \ \geq  \ \frac{1}{16\pi^2}\ \![(\mathbb{J}_r)_{ii}(\mathcal{X})]\, ,
\label{app.c.65}
\end{eqnarray}
where in the last inequality we have used the fact that $q^{1/(1-q)} \geq 1/4$ for $q \in [1/2,1]$ and that $[\Gamma(D/2 +1)]^{2/D} \geq \pi/4$.
As a final step we employ Eq.~(\ref{app.B.58}) and (\ref{app.c.65}) to write
\begin{eqnarray}
N_{q}(\mathcal{Y})\ \geq \ \frac{1}{16\pi^2 D} \ \! \mbox{Tr}({\mathbb{J}}_r(\mathcal{X}))\ \geq \ \frac{1}{16\pi^2 } \ \! [\det({\mathbb{J}}_r(\mathcal{X}))]^{1/D}\, ,
\end{eqnarray}
which completes the proof of the generalized Stam's inequality.

%\section{}
%All appendix sections must be cited in the main text. In the appendixes, Figures, Tables, etc. should be labeled starting with `A', e.g., Figure A1, Figure A2, etc.

%%%%%%%%%%%%%%%%%%%%%%%%%%%%%%%%%%%%%%%%%%
\reftitle{References}

% Please provide either the correct journal abbreviation (e.g. according to the “List of Title Word Abbreviations” http://www.issn.org/services/online-services/access-to-the-ltwa/) or the full name of the journal.
% Citations and References in Supplementary files are permitted provided that they also appear in the reference list here.

%=====================================
% References, variant A: external bibliography
%=====================================
%\externalbibliography{yes}
%\bibliography{your_external_BibTeX_file}

%=====================================
% References, variant B: internal bibliography
%=====================================

%%%%%%%%%%%%%%%%%%%%%%%%%%%%%%%%%%%%%%%%%%
\end{document}